\documentclass[%
 aip,
 amsmath,amssymb,
 reprint,%
]{revtex4-1}

\usepackage{amsmath}
\usepackage{graphicx}
\usepackage{xspace}
\usepackage{multirow}
\usepackage{latexsym}
\usepackage{subfig}
\usepackage{braket}
\usepackage{cleveref}
\usepackage{threeparttable}
\usepackage{enumerate}
\usepackage{mathrsfs}
\usepackage{enumitem}
\usepackage{color}
\usepackage[version=3]{mhchem}
\usepackage{caption,setspace}

\allowdisplaybreaks
\usepackage{array}
\newcolumntype{L}[1]{>{\raggedright\let\newline\\\arraybackslash\hspace{0pt}}m{#1}}
\newcolumntype{C}[1]{>{\centering\let\newline\\\arraybackslash\hspace{0pt}}m{#1}}
\newcolumntype{R}[1]{>{\raggedleft\let\newline\\\arraybackslash\hspace{0pt}}m{#1}}

\crefname{figure}{Figure}{Figures}
\crefname{table}{Table}{Tables}
\crefname{equation}{Eq.}{Eqs.}
\crefname{section}{Section}{Sections}

\newcommand*{\degree}{\ensuremath{^\circ}\xspace}
\newcommand{\h}[2]{h_{{#1}}^{{#2}}}
\newcommand{\f}[2]{f_{{#1}}^{{#2}}}
\renewcommand{\d}[2]{\delta_{{#1}}^{{#2}}}
\renewcommand{\v}[2]{{v}_{{#1}}^{{#2}}}
\renewcommand{\t}[2]{{t}_{{#1}}^{{#2}}}
\renewcommand{\c}[1]{a^\dagger_{#1}}
\renewcommand{\a}[1]{a^{\ }_{#1}}
\newcommand{\pdm}[2]{\gamma_{{#1}}^{{#2}}}

\newcommand{\e}[1]{\ensuremath{\varepsilon_{#1}}}

\newcommand*{\mae}{$\Delta_{\mathrm{MAE}}$\xspace}
\newcommand*{\std}{$\Delta_{\mathrm{STD}}$\xspace}
\newcommand*{\eh}{\ensuremath{E_\mathrm{h}}\xspace}
\newcommand{\angstrom}{\mbox{\normalfont\AA}}

\begin{document}

\author{Alexander~Yu.~Sokolov}
\email{sokolov.8@osu.edu}
\affiliation{Department of Chemistry and Biochemistry, The Ohio State University, Columbus, Ohio 43210, USA}
\title{Multi-reference algebraic diagrammatic construction theory for excited states: General formulation and first-order implementation}

\begin{abstract}
We present a multi-reference generalization of the algebraic diagrammatic construction theory (ADC) [J.\@ Schirmer, Phys.\@ Rev.\@ A {\bf 26}, 2395 (1982)] for excited electronic states. The resulting multi-reference ADC approach (MR-ADC) can be efficiently and reliably applied to systems, which exhibit strong electron correlation in the ground or excited electronic states. In contrast to conventional multi-reference perturbation theories, MR-ADC describes electronic transitions involving all orbitals (core, active, and external) and enables efficient computation of spectroscopic properties, such as transition amplitudes and spectral densities. 
Our derivation of MR-ADC is based on the effective Liouvillean formalism of Mukherjee and Kutzelnigg [D. Mukherjee, W. Kutzelnigg, in {\it Many-Body Methods in Quantum Chemistry} (1989), pp. 257--274], which we generalize to multi-determinant reference states. 
We discuss a general formulation of MR-ADC, perform its perturbative analysis, and present an implementation of the first-order MR-ADC approximation, termed MR-ADC(1), as a first step in defining the MR-ADC hierarchy of methods. We show results of MR-ADC(1) for the excitation energies of the Be atom, an avoided crossing in \ce{LiF}, doubly excited states in \ce{C2}, and outline directions for our future developments.
\end{abstract}

\titlepage

\maketitle

\section{Introduction}
\label{sec:intro}

Accurate description of excited electronic states and strong electron correlation are among the greatest challenges in modern quantum chemistry. Theoretical approaches for excited states can be divided into the wavefunction and propagator (or linear-response) categories. The wavefunction methods compute properties of each electronic state individually, from the wavefunctions and energies obtained by solving the Schr\"odinger equation.\cite{Szabo:1982,Helgaker:2000} In contrast, the propagator methods directly compute the energy differences and the transition amplitudes between electronic states, from the poles and residues of the approximate propagators.\cite{Fetter2003,Dickhoff2008} Additionally, the wavefunction and propagator methods can be classified as single-reference or multi-reference, based on their ability to describe strong electron correlation. 

Among the single-reference approaches, many wavefunction and propagator methods have been developed and their strengths and weaknesses have been well documented. The wavefunction 
theories\cite{Nakatsuji:1978p2053,Nakatsuji:1979p329,Nooijen:1997p6441,Nooijen:1997p6812,Sherrill:1999p143,Geertsen:1989p57,Comeau:1993p414,Stanton:1993p7029,Krylov:2008p433}
offer a hierarchy of approximations that can be used to compute accurate excitation energies for small molecules. Meanwhile, the propagator 
methods\cite{Goscinski:1980p385,Weiner:1980p1109,Prasad:1985p1287,Datta:1993p3632,Lowdin:1970p231,Nielsen:1980p6238,Sangfelt:1984p3976,Bak:2000p4173,Mukherjee:1989p257,Nooijen:1992p55,Nooijen:1993p15,Nooijen:1995p1681,Moszynski:2005p1109,Korona:2010p14977,Kowalski:2014p094102,Schirmer:1982p2395,Schirmer:1983p1237,Schirmer:1991p4647,Mertins:1996p2140,Schirmer:2004p11449,Dreuw:2014p82,Liu:2018p244110} 
provide a direct access to important spectroscopic properties, such as transition amplitudes and spectral densities, often at a lower computational cost. Although the two types of methods significantly differ in their theoretical foundation, it has been demonstrated that the propagator methods have a close connection to the wavefunction theories formulated using effective Hamiltonians.\cite{Mukherjee:1989p257,Tarantelli:1989p233,Nooijen:1992p55,Nooijen:1993p15,Nooijen:1995p1681,Moszynski:2005p1109,Korona:2010p14977,Kowalski:2014p094102,McClain:2016p235139,BhaskaranNair:2016p144101,Lange:2018p4224,Berkelbach:2018p041103} For example, equation-of-motion coupled cluster theory (EOM-CC), which is widely regarded as a wavefunction approach,\cite{Geertsen:1989p57,Comeau:1993p414,Stanton:1993p7029,Krylov:2008p433} yields excitation energies that are equivalent to those of linear-response coupled cluster theory.\cite{Sekino:1984p255,Koch:1990p3345,Koch:1990p3333} Connections between EOM-CC and Green's function coupled cluster formalism\cite{Nooijen:1992p55,Nooijen:1993p15,Nooijen:1995p1681,Moszynski:2005p1109,Korona:2010p14977,Kowalski:2014p094102,McClain:2016p235139,BhaskaranNair:2016p144101} have been a matter of extensive study. A recent work established connections between approximate EOM-CC and the random phase approximation.\cite{Berkelbach:2018p041103}

For excited states of strongly correlated systems, a wide range of multi-reference approaches have been explored.\cite{Werner:1980p2342,Werner:1981p5794,Knowles:1985p259,Wolinski:1987p225,Hirao:1992p374,Werner:1996p645,Finley:1998p299,Andersson:1990p5483,Andersson:1992p1218,Angeli:2001p10252,Angeli:2001p297,Angeli:2004p4043,Banerjee:1979p1209,Yeager:1979p77,Dalgaard:1980p816,Yeager:1984p85,Graham:1991p2884,Yeager:1992p133,Nichols:1998p293,Khrustov:2002p507,Chattopadhyay:2000p7939,Chattopadhyay:2007p1787,Jagau:2012p044116,Samanta:2014p134108,Datta:2012p204107,Nooijen:2014p081102,Huntington:2015p194111,Aoto:2016p074103} These methods usually combine a high-level description of strong electron correlation in a small subset of near-degenerate (active) orbitals and a lower-level description of weaker (dynamic) correlation in the remaining (core and external) orbitals. Here, it is important to distinguish strong electron correlation in the ground and excited electronic states. In the wavefunction-based multi-reference theories, such as complete active-space self-consistent field (CASSCF) \cite{Werner:1980p2342,Werner:1981p5794,Knowles:1985p259} or multi-reference perturbation (MRPT) methods,\cite{Wolinski:1987p225,Hirao:1992p374,Werner:1996p645,Finley:1998p299,Andersson:1990p5483,Andersson:1992p1218,Angeli:2001p10252,Angeli:2001p297,Angeli:2004p4043} strong correlation in the ground and excited states is described by constructing a different multi-reference wavefunction for each state. While these approaches provide an unbiased description of strong correlation in various states, they often lack accurate treatment of dynamic correlation or become computationally expensive for large active spaces, basis sets, or many electronic states. For this reason, the MRPT methods have become particularly popular as they combine the description of dynamic correlation with a relatively low computational cost and can be used to compute reasonably accurate excitation energies for large systems. Recent developments enable simulations of excited states using MRPT with large basis sets and many (more than 20) active orbitals.\cite{Kurashige:2011p094104,Kurashige:2014p174111,Guo:2016p1583,Sharma:2017p488,Yanai:2017p4829,Freitag:2017p451,Sokolov:2017p244102} Although the MRPT methods are computationally efficient, their application to other excited-state properties is hindered by the complexity of their analytic derivative expressions.\cite{MacLeod:2015p051103} Another limitation is that the MRPT methods can only simulate electronic states originating from the transitions between active orbitals, and thus cannot be used to compute large spectral regions where excitations into the higher-energy (external) or from the lower-energy (core) orbitals may be important.

A different strategy for describing strong correlation in various electronic states is offered by the multi-reference propagator,\cite{Banerjee:1979p1209,Yeager:1979p77,Dalgaard:1980p816,Yeager:1984p85,Graham:1991p2884,Yeager:1992p133,Nichols:1998p293,Khrustov:2002p507} linear-response,\cite{Chattopadhyay:2000p7939,Chattopadhyay:2007p1787,Jagau:2012p044116,Samanta:2014p134108} and equation-of-motion approaches.\cite{Datta:2012p204107,Nooijen:2014p081102,Huntington:2015p194111} In these methods, a single electronic state (usually, the ground state) is chosen as the reference (or ``parent'') state and the remaining (``target'') states are expressed with respect to this reference. Here, the parent state is used to obtain information about the dynamic correlation, and the correlation in the target states is assumed to be similar. This group of methods can be used to compute excitation energies for many electronic states simultaneously, including excitations outside of the active space, provided that the parent state is suitable for describing the target states and that the reference and excited states do not cross. For example, multi-reference equation-of-motion coupled cluster theory (MR-EOM-CC) developed by Nooijen and co-workers\cite{Datta:2012p204107,Nooijen:2014p081102,Huntington:2015p194111} incorporates single excitations involving core and external orbitals and can describe many (100's) electronic states simultaneously. This method is, however, not size-extensive and has a relatively higher computational cost compared to MRPT. Alternatively, multi-reference propagator methods\cite{Banerjee:1979p1209,Yeager:1979p77,Dalgaard:1980p816,Yeager:1984p85,Graham:1991p2884,Yeager:1992p133,Nichols:1998p293,Khrustov:2002p507} are computationally more efficient than MR-EOM-CC, but lack accurate description of the two-electron dynamic correlation effects outside of the active space. Additionally, most of the multi-reference propagator, linear-response, and equation-of-motion theories are based the non-Hermitian eigenvalue problems that can be sensitive to various instabilities common in multi-reference theories, giving rise to unphysical excitation energies.

A special class of the single-reference propagator methods is the algebraic diagrammatic construction theory (ADC).\cite{Schirmer:1982p2395,Schirmer:1983p1237,Schirmer:1991p4647,Mertins:1996p2140,Schirmer:2004p11449,Dreuw:2014p82} Among the attractive properties of ADC are relatively low computational cost, Hermitian eigenvalue problem, and efficient access to excited-state properties. Although originally formulated from a perturbative diagrammatic analysis of the time-dependent polarization propagator, ADC has been later rederived in the time-independent context by starting with a ground-state wavefunction from M\o ller-Plesset perturbation theory using the so-called intermediate-state representation approach.\cite{Schirmer:1991p4647,Mertins:1996p2140,Schirmer:2004p11449} An alternative derivation of ADC has been suggested by Mukherjee and Kutzelnigg within the framework of the effective Liouvillean formalism.\cite{Mukherjee:1989p257} This framework has a close connection to consistent propagator theory developed earlier by Prasad et al.,\cite{Prasad:1985p1287} the time-independent Fock-space Green's function theory,\cite{Kutzelnigg:1998p5578} and single-reference unitary coupled cluster theory (UCC).\cite{Kutzelnigg:1982p3081,Bartlett:1989p133,Watts:1989p359} Recently, it has been demonstrated that ADC emerges as an approximation in the linear-response UCC theory\cite{Kats:2011p062503,Walz:2012p052519} and self-consistent UCC-based polarization propagator theory.\cite{Liu:2018p244110}

One of the main limitations of ADC is its inherently single-reference nature, which prevents applications to systems with open-shell and multi-reference character in the ground or excited electronic states. A spin-flip version of ADC has been shown to provide accurate results for some multi-reference systems that possess a single-reference triplet ground state.\cite{Lefrancois:2015p124107,Lefrancois:2017p4436} However, a general multi-reference formulation of ADC has not been developed, to the best of our knowledge. 

In this work, we present a multi-reference formulation of ADC (MR-ADC) for excited states of strongly correlated systems. We demonstrate that such MR-ADC formulation can be achieved by combining the effective Liouvillean formalism of Mukherjee and Kutzelnigg\cite{Mukherjee:1989p257} with multi-reference perturbation theory and can be considered as a natural generalization of the conventional ADC theory for the multi-configurational reference wavefunctions. In \cref{sec:theory_background}, we give a brief overview of the effective Liouvillean theory and outline the derivation of single-reference ADC using this approach. Next, in \cref{sec:theory_mr_adc}, we describe a general formulation of MR-ADC for the polarization propagator and provide a recipe for constructing MR-ADC approximations at each order in perturbation theory. As an example, in \cref{sec:implementation}, we present an implementation of the first-order MR-ADC approximation (MR-ADC(1)). We outline computational details in \cref{sec:comp_details}, and benchmark MR-ADC(1) for small systems in \cref{sec:results}. Finally, in \cref{sec:conclusions}, we present our conclusions and outline plans for future developments.

\section{Theory: Background}
\label{sec:theory_background}

\subsection{Propagators and the effective Liouvillean theory}
\label{sec:theory_background_liouvillean}

All ADC schemes have a close connection to the propagator theory. A general form of the retarded frequency-dependent propagator can be written as:\cite{Fetter2003,Dickhoff2008}
\begin{align}
	\label{eq:g_munu}
	G_{\mu\nu}(\omega) 
	& = G_{\mu\nu}^+(\omega) \pm G_{\mu\nu}^-(\omega) \notag\\
	& =  \bra{\Psi}q_\mu(\omega - H + E)^{-1}q^\dag_\nu\ket{\Psi} \notag \\
	&\pm \bra{\Psi}q^\dag_\nu(\omega + H - E)^{-1}q_\mu\ket{\Psi} 
\end{align}
where $G_{\mu\nu}^+(\omega)$ and $G_{\mu\nu}^-(\omega)$ are the forward and backward components of the propagator and the wavefunction $\ket{\Psi}$ is an eigenstate of the Hamiltonian $H$ with an eigenvalue $E$. The frequency can be defined as $\omega \equiv \omega' + i\eta$, where $\omega'$ is the real component of $\omega$ and $i\eta$ is an infinitesimal imaginary number. The operators $q^\dag_\nu$ depend on the propagator of interest. For example, for the polarization propagator, $q^\dag_\nu = \c{p}\a{q} - \braket{\Psi|\c{p}\a{q}|\Psi}$, where $\c{p}$ and $\a{p}$ are the usual creation and annihilation operators. The $+$ or $-$ signs are chosen if $q^\dag_\nu$ are the products of an odd or even number of creation/annihilation operators, respectively. 

\cref{eq:g_munu} can be written in a more compact form
\begin{align}
	\label{eq:g_munu_2}
	G_{\mu\nu}(\omega) 
	& =  \bra{\Psi}[q_\mu,(\omega - \mathcal{H})^{-1}q^\dag_\nu]_{\pm}\ket{\Psi}
\end{align}
where $[\ldots]_\pm$ denotes anti-commutator or commutator (corresponding to $+$ or $-$ sign in \cref{eq:g_munu}, respectively) and $\mathcal{H}$ is the Liouvillean superoperator\cite{Lowdin:1985p285} with the following property: $\mathcal{H}A = [H,A] = H A - A H$, where $A$ is an arbitrary operator. Introducing the binary product of two operators\cite{Lowdin:1968p867}
\begin{align}
	(A|B) = \braket{\Psi|[A,B^\dag]_\pm|\Psi}
\end{align}
\cref{eq:g_munu_2} can be expressed in a matrix form:
\begin{align}
\label{eq:inner_proj}
	\mathbf{G}(\omega) 
	& = \mathbf{T^{}_{X}} \mathbf{A_{X}^{-1}}(\omega) \mathbf{T_{X}^\dag}
\end{align}
where the $\mathbf{T^{}_{X}}$ and $\mathbf{A^{}_{X}}(\omega)$ matrices are defined as:
\begin{align}
	\label{eq:T_general}
	\mathbf{T^{}_{X}} &= 
	\begin{bmatrix}
		(\mathbf{q} | \mathbf{X_{+}}) & (\mathbf{q} | \mathbf{X_{-}^\dag})
	\end{bmatrix} \\
	\label{eq:A_general}
	\mathbf{A^{}_{X}}(\omega) &= 
	\begin{bmatrix}
		(\mathbf{X_{+}}|\omega-\mathcal{H}|\mathbf{X_{+}}) & (\mathbf{X_{+}}|\omega - \mathcal{H}|\mathbf{X^\dag_{-}}) \\
		(\mathbf{X^\dag_{-}}|\omega - \mathcal{H}|\mathbf{X_{+}}) & (\mathbf{X^\dag_{-}}|\omega - \mathcal{H}|\mathbf{X^\dag_{-}})
	\end{bmatrix}
\end{align}
Here, $\mathbf{q}$ denotes the full set of operators $q_\mu$ and $\mathbf{X^\dag_{+}}$ ($\mathbf{X^\dag_{-}}$) is the projection operator manifold\cite{Lowdin:1965pA357,Goscinski:1970p573,Manne:1977p470,Dalgaard:1979p169} for $\mathbf{G_{+}}$ ($\mathbf{G_{-}}$) with elements $X^\dag_{+\mu}$ ($X^\dag_{-\mu}$). \cref{eq:inner_proj} can be used to compute $\mathbf{G}(\omega)$ exactly, provided that the projection manifolds $\mathbf{X^\dag_{\pm}}$ are complete. Importantly, even for complete $\mathbf{X^\dag_{\pm}}$, the matrix $\mathbf{A^{}_{X}}(\omega)$ in general contains elements that couple $\mathbf{G_{+}}$ and $\mathbf{G_{-}}$,\cite{Mukherjee:1989p257} i.e. \mbox{$(\mathbf{X_{+}}|\omega - \mathcal{H}|\mathbf{X^\dag_{-}}) \ne 0$}. In order to ensure that this coupling is zero, the operators $X^\dag_{\pm\mu}$ must fulfill the ``vacuum annihilation condition'' (VAC):\cite{Goscinski:1980p385,Weiner:1980p1109,Prasad:1985p1287,Datta:1993p3632} 
\begin{align}
\label{eq:vac}
X_{\pm\mu}\ket{\Psi} = 0 
\end{align}
However, in practice, satisfying VAC can be very difficult, due to a rather complicated form of \cref{eq:vac} for a general correlated ground state $\ket{\Psi}$.

A procedure to construct (incomplete) operator manifolds that satisfy VAC even for approximate correlated wavefunctions was first proposed by Prasad et al.\cite{Prasad:1985p1287} and was later developed by Mukherjee and Kutzelnigg within the framework of the effective Liouvillean theory.\cite{Mukherjee:1989p257} In the first step of this approach, the ground-state wavefunction is expressed using a unitary cluster expansion:
\begin{align}
	\label{eq:ucc_wfn}
	\ket{\Psi} &= e^{A} \ket{\Phi} , \qquad A = T - T^\dag
\end{align}
where $\ket{\Phi}$ is a single-determinant reference state and $T$ is an excitation operator. In the second step, a new operator manifold $h^\dag_{\pm \mu}$ is defined that satisfies VAC with respect to the {\it model} state: $h_{\pm \mu}\ket{\Phi} = 0$. Due to a simple structure of $\ket{\Phi}$, the form of $h^\dag_{\pm \mu}$ is also rather simple, they can be expressed as products of creation and annihilation operators with unoccupied and occupied orbital labels, respectively. Finally, the operators $h^\dag_{\pm \mu}$ are used to define another operator manifold $\tilde{X}^\dag_{\pm\mu}$ that fulfills VAC for the {\it correlated} ground state:
\begin{align}
	\tilde{X}^\dag_{\pm\mu} &= e^{A} h^\dag_{\pm \mu} e^{-A} \\
	\label{eq:consistent_op}
	\tilde{X}_{\pm\mu} \ket{\Psi} &= e^{A} h_{\pm \mu} \ket{\Phi} = 0 
\end{align}
Replacing $\mathbf{X_{\pm}^\dag}$ in \cref{eq:T_general,eq:A_general} by $\mathbf{\tilde{X}_{\pm}^\dag}$ decouples the forward and backward components of the propagator, which now takes the form:
\begin{align}
	\label{eq:G_decoupled}
	\mathbf{G}(\omega) 
	& = \mathbf{T^{}_{\tilde{X}}} \mathbf{A_{\tilde{X}}^{-1}}(\omega) \mathbf{T_{\tilde{X}}^\dag} \\
	\label{eq:T_decoupled}
	\mathbf{T_{\tilde{X}}} &= 
	\begin{bmatrix}
		\{\mathbf{\tilde{q}} | \mathbf{h_{+}}\} & \{\mathbf{\tilde{q}} | \mathbf{h_{-}^\dag}\}
	\end{bmatrix} \\
	\label{eq:A_decoupled}
	\mathbf{A_{\tilde{X}}} &= 
	\begin{bmatrix}
		\{\mathbf{h_{+}}|\omega-\tilde{\mathcal{H}}|\mathbf{h_{+}}\} & 0 \\
		0 & \{\mathbf{h^\dag_{-}}|\omega - \tilde{\mathcal{H}}|\mathbf{h^\dag_{-}}\}
	\end{bmatrix}
\end{align}
where $\mathbf{h_{\pm}^\dag}$ is a collection of $h^\dag_{\pm \mu}$, $\mathbf{\tilde{q}}$ is a set of transformed operators $\tilde{q}_\mu^\dag = e^{-A} q_\mu^\dag e^{A}$, $\tilde{\mathcal{H}}$ is a superoperator corresponding to the effective Hamiltonian $\tilde{H} = e^{-A} H e^{A}$, and a new notation for the binary product of two operators with respect to the model state $\ket{\Phi}$ is introduced:
\begin{align}
	\{A|B\} &= \braket{\Phi|[A,B^\dag]_\pm|\Phi}
\end{align}
Importantly, the operators $\tilde{X}^\dag_{\pm\mu}$ fulfill VAC and the decoupling in \cref{eq:A_decoupled} is achieved even when the cluster operator $A$ is truncated at a low excitation rank, provided that the following condition is satisfied:
\begin{align}
	\label{eq:H_eff_proj}
	\braket{\Phi|h_{+\mu}h_{-\nu}e^{-A} H e^{A}|\Phi} = 0
\end{align}
i.e. the corresponding projection of the effective Hamiltonian $\tilde{H}$ vanishes. In order to satisfy \cref{eq:H_eff_proj}, the excitation rank of the operator $A$ must not be lower than the total deexcitation rank of the operator $h_{+\mu}h_{-\nu}$. For example, if $h_{+\mu}$ and $h_{-\nu}$ are both single-deexcitation operators, the operator $A$ in \cref{eq:H_eff_proj} must include up to two-body terms. 

We note that, due to the unitary nature of the wave operator $e^{A}$, in the effective Liouvillean approach expression for $\tilde{H} = e^{-A} H e^{A}$ does not terminate. An alternative formalism based on the extended coupled cluster parametrization of the ground-state wavefunction has been explored by Datta et al.\cite{Datta:1993p3632} This approach uses a non-Hermitian effective Hamiltonian, but gives rise to expressions with a finite number of terms.

\subsection{Single-reference ADC from the effective Liouvillean theory}
\label{sec:theory_background_sr_adc}

In this section, we briefly describe how the effective Liouvillean theory can be used to derive approximations of single-reference ADC (SR-ADC). The conventional derivation of SR-ADC uses the so-called intermediate state representation (ISR) approach.\cite{Schirmer:1991p4647,Mertins:1996p2140,Schirmer:2004p11449} While it has been useful for developing new SR-ADC methods,\cite{Schirmer:1998p4734,Trofimov:2005p144115,Starcke:2009p024104,Knippenberg:2012p064107,Pernpointner:2014p084108,Lefrancois:2015p124107} the ISR approach does not admit a straightforward generalization for multi-reference wavefunctions. In contrast, the effective Liouvillean approach has a close connection to unitary coupled cluster theory and many-body perturbation theory, making such generalization possible. For the polarization propagator, both derivations lead to the identical equations for the excitation energies and transition amplitudes of the SR-ADC approximations. 
A similar approach developed by Liu et al.\cite{Liu:2018p244110}\@ has been recently used to derive the ADC(3) approximation for the polarization propagator and its self-consistent variant.

Starting with a single-determinant reference wavefunction $\ket{\Phi}$, the electronic Hamiltonian
\begin{align}
	\label{eq:hamiltonian}
	H = \sum_{pq} \h{p}{q} \c{p}\a{q} + \frac{1}{4} \sum_{pqrs} \v{pq}{rs} \c{p}\c{q}\a{s}\a{r} \
\end{align}
can be expressed in the normal-ordered form
\begin{align}
	\label{eq:h_no_sr}
	H = E_0 + \sum_{pq} (f_0)_p^q \{ \c{p}\a{q} \} + \frac{1}{4} \sum_{pqrs} \v{pq}{rs} \{ \c{p}\c{q}\a{s}\a{r} \}
\end{align}
where $E_0 = \braket{\Phi|H|\Phi}$, $\h{p}{q} = \braket{p|h|q}$, $\v{pq}{rs} = \braket{pq||rs}$, and $(f_0)_p^q = \h{p}{q} + \sum_{i}^{occ} \v{pi}{qi}$ are, respectively, the reference energy, the one-electron integrals, the antisymmetrized two-electron integrals, and the canonical Fock matrix. Notation $\{\ldots\}$ indicates that the creation and annihilation operators are normal-ordered with respect to $\ket{\Phi}$. Indices $p,q,r,s$ run over all spin-orbitals in a finite one-electron basis set. 

To derive the SR-ADC approximations, we partition the Hamiltonian into the zeroth-order part
\begin{align}
	\label{eq:h_zero_sr}
	H^{(0)} = E_0 + \sum_{p} \e{p} \{ \c{p}\a{p} \}
\end{align}
and the perturbation $V = H - H^{(0)}$, where the Fock matrix is assumed to be diagonal: $(f_0)_p^q = \e{p}\d{p}{q}$. This leads to a perturbative expansion for the wavefunction
\begin{align}
	\ket{\Psi} &= e^{A^{(0)} + A^{(1)} + \ldots + A^{(n)} + \ldots} \ket{\Phi} 
\end{align}
and the propagator
\begin{align}
	\label{eq:g_pt_series}
	\mathbf{G}(\omega) & = \mathbf{G}^{(0)}(\omega) + \mathbf{G}^{(1)}(\omega) + \ldots + \mathbf{G}^{(n)}(\omega) + \ldots
\end{align}
For the polarization propagator, truncating expansion for $\mathbf{G}(\omega)$ at the $n$-th order leads to the equations for the single-reference ADC(n) approximation.\cite{Schirmer:1982p2395} 

If $\mathbf{G}(\omega)$ is expressed in the form of \cref{eq:G_decoupled}, the forward and backward components of $\mathbf{G}(\omega)$ are decoupled and thus can be considered separately. 
As an example, we consider the $n$th-order contribution to $\mathbf{G}_{+}(\omega)$, which can be written as:
\begin{align}
	\label{eq:Gn_matrix}
	\mathbf{G}_{+}^{(n)}(\omega) & = \mathbf{T}^{(n)}_{+} \mathbf{A}_{+}^{\mathbf{-1}\,(n)}(\omega) \mathbf{T}_{+}^{(n)\dag} 
\end{align}
where the subscript $\mathbf{\tilde{X}}$ that appears in \cref{eq:G_decoupled} is omitted for clarity. Here, $\mathbf{T}^{(n)}_{+}$ is the $n$-th-order contribution to the matrix of the SR-ADC effective transition moments
\begin{align}
	\label{eq:T_matrix}
	\mathbf{T}^{(n)}_{+} & = \sum_{kl}^{k+l=n} \{\mathbf{\tilde{q}}^{(k)} | \mathbf{h}_{+}^{(l)}\}
\end{align}
The matrix $\mathbf{A}_{+}^{(n)}(\omega)$ can be expressed as:
\begin{align}
	\label{eq:A_matrix}
	\mathbf{A}_{+}^{(n)}(\omega) & =  \omega \mathbf{S}_{+}^{(n)} - \mathbf{M}_{+}^{(n)} \\
	\label{eq:S_matrix}
	\mathbf{S}_{+}^{(n)} & = \sum_{kl}^{k+l= n} \{\mathbf{h}_{+}^{(k)}|\mathbf{h}_{+}^{(l)}\} \\
	\label{eq:M_matrix}
	\mathbf{M}_{+}^{(n)} & = \sum_{klm}^{k+l+m= n} \{\mathbf{h}_{+}^{(k)}|\tilde{\mathcal{H}}^{(l)}|\mathbf{h}_{+}^{(m)}\}
\end{align}
where $\mathbf{M}_{+}^{(n)}$ and $\mathbf{S}_{+}^{(n)}$ are the $n$th-order contributions to the effective Liouvillean and overlap matrices, respectively. The former matrix is usually called the effective Hamiltonian matrix in the SR-ADC literature,\cite{Dreuw:2014p82} it contains information about the excitation energies. 

To determine $\mathbf{T}^{(n)}_{+}$, $\mathbf{S}_{+}^{(n)}$, and $\mathbf{M}_{+}^{(n)}$, equations for $\mathbf{\tilde{q}}^{(k)\dag}$, $\tilde{H}^{(k)}$, and $\mathbf{h}_{+}^{(k)\dag}$ are derived at each perturbative order. Expanding $\mathbf{\tilde{q}^\dag}$ and $\tilde{H}$ using the Baker--Campbell--Hausdorff (BCH) formula
\begin{align}
	\label{eq:q_bch}
	\tilde{q}^{\dag}_\mu &= q^\dag_\mu + [q^\dag_\mu, A] + \frac{1}{2!} [[q^\dag_\mu, A],A] + \ldots \\
	\label{eq:H_bch}
	\tilde{H} &= H + [H, A] + \frac{1}{2!} [[H, A], A] + \ldots 
\end{align}
and collecting the $k$-th-order terms on both sides of the equations, allows to obtain $\mathbf{\tilde{q}}^{(k)\dag}$ and $\tilde{H}^{(k)}$. Once $\mathbf{\tilde{q}^\dag}$ is determined up to the $k$-th order, the sum of its contributions can be expressed in the following form:
\begin{align}
	\label{eq:determine_h}
	\tilde{q}_\mu^{(0)\dag} + \tilde{q}_\mu^{(1)\dag} + \ldots + \tilde{q}_\mu^{(k)\dag} = \sum_{\nu} h^{m,n \, \dag }_{\pm\nu} d_{\mu\nu}^{(k)}
\end{align}
The r.h.s.\@ of \cref{eq:determine_h} is a sum of all operators $h^{m,n \, \dag}_{\pm\nu}$ that appear in the BCH expansion \eqref{eq:q_bch} up to the $k$-th order, $d_{\mu\nu}^{(k)}$ are the linear coefficients. Each operator $h^{m,n \, \dag}_{\pm\nu}$ is classified by its particle-hole rank, where $m$ and $n$ are the numbers of particle and hole labels, respectively. \cref{eq:determine_h} defines the operators $h^{m,n \, \dag}_{\pm\nu}$ that compose the operator manifolds $\mathbf{h}_{\pm}^{(k)\dag}$ up to the $k$-th order. For example, $\mathbf{h}_{+}^{(0)\dag}$ consists of the operators that have the same particle-hole rank as $\mathbf{q^\dag}$. The operators $\mathbf{h}_{+}^{(1)\dag}$ include all new operators generated by the first commutator in \cref{eq:q_bch} that are not included in $\mathbf{h}_{+}^{(0)\dag}$. This procedure can be repeated to obtain $\mathbf{h}_{+}^{(k)\dag}$ at an arbitrary order.

Finally, the SR-ADC excitation energies are determined by truncating the perturbation expansion of $\mathbf{M}_{+}$ and $\mathbf{S}_{+}$ at the $n$-th order and solving the generalized eigenvalue problem:
\begin{align}
	\label{eq:adc_eig_problem}
	&\mathbf{M}_{+} \mathbf{Y} = \mathbf{S}_{+} \mathbf{Y} \boldsymbol{\Omega} \\
	\label{eq:adc_eig_problem_2}
	\mathbf{M}_{+} &\approx \mathbf{M}_{+}^{(0)} + \ldots + \mathbf{M}_{+}^{(n)} \\
	\label{eq:adc_eig_problem_3}
	\mathbf{S}_{+} &\approx \mathbf{S}_{+}^{(0)} + \ldots + \mathbf{S}_{+}^{(n)} 
\end{align}
where $\boldsymbol{\Omega}$ is a diagonal matrix of excitation energies and $\mathbf{Y}$ are the eigenvectors. To compute $\mathbf{M}_{+}^{(k)}$, the amplitudes of the operators $T$ and $T^\dag$ in \cref{eq:ucc_wfn} need to be determined. Those are obtained by projecting the effective Hamiltonian according to \cref{eq:H_eff_proj}. The details of this procedure will be demonstrated in \cref{sec:theory_mr_adc}, where we will discuss the derivation of multi-reference ADC.

As pointed out to us by one of the Reviewers, an alternative hierarchy for constructing approximations for the propagator $\mathbf{G}(\omega)$ is to consider the perturbation series for the forward and backward components of $\mathbf{G}(\omega)$ separately from the outset. This strategy has a close connection to the intermediate state representation approach and does not require the use of operator manifolds that satisfy VAC, but is less attractive from a perturbative standpoint. 

\section{Theory: Multi-reference Algebraic Diagrammatic Construction (MR-ADC)}
\label{sec:theory_mr_adc}

\subsection{General aspects of MR-ADC for the polarization propagator}
\label{sec:theory_mr_adc_general_aspects}

\begin{figure}[t!]
	\includegraphics[width=0.48\textwidth]{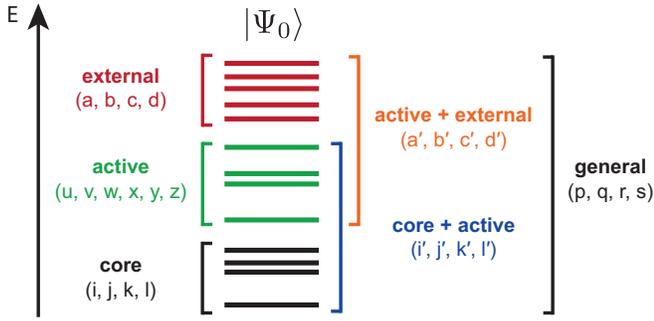}
	\captionsetup{justification=raggedright,singlelinecheck=false}
	\caption{Orbital energy diagram showing the index convention used in this work.}
	\label{fig:mo_diagram}
\end{figure}                                                                                                                                                                             

Here, we present a general formulation of the multi-reference ADC theory for the polarization propagator (MR-ADC). We start by dividing the spin-orbitals into three sets: core, active, and external. \cref{fig:mo_diagram} shows the orbital spaces and the orbital index notation used in this work. We now assume that we have solved the complete active-space self-consistent field (CASSCF) variational problem and computed the reference wavefunction $\ket{\Psi_0}$ for the ground state. In addition to $\ket{\Psi_0}$, our model space contains the excited-state wavefunctions $\ket{\Psi_I}$ ($I>0$), which we obtain from the complete active-space configuration interaction computation (CASCI) using the ground-state CASSCF orbitals. We refer to this procedure of constructing the model space as CASCI/CASSCF.

To guide our development of MR-ADC further, we introduce two requirements: 
\begin{itemize}
\item
{\it Requirement 1}. At each order of perturbation theory $n$, the $n$-th-order MR-ADC approximation must reduce to the $n$-th-order SR-ADC approximation [ADC(n)] in the limit of the single-determinant $\ket{\Psi_0}$ and zero active orbitals.
\item
{\it Requirement 2}. The zeroth-order \mbox{MR-ADC} approximation must yield exactly the CASCI/CASSCF excitation energies and transition amplitudes between the ground- ($\ket{\Psi_0}$) and excited-state ($\ket{\Psi_I}$, $I > 0$) model wavefunctions in the active space.
\end{itemize}
These two requirements make sure that MR-ADC produces the SR-ADC and the exact (i.e., full configuration interaction) excitation energies and transition amplitudes in the two limits, respectively: (i) zero active orbitals (requirement 1), and (ii) all orbitals are active (requirement 2). 

To satisfy the requirement 1, we derive MR-ADC using the effective Liouvillean approach described in \cref{sec:theory_background} that we generalize for multi-determinant reference wavefunctions. Thus, we consider approximations to the ground-state correlated wavefunction in the form:
\begin{align}
	\label{eq:mr_adc_wfn}
	\ket{\Psi} &= e^{A} \ket{\Psi_0} = e^{T - T^\dag} \ket{\Psi_0} , \quad T = \sum_{k=1}^N T_k  \\
	\label{eq:mr_adc_t_amplitudes}
	T_k &= \frac{1}{(k!)^2} {\sum_{i'j'a'b'\ldots}} t_{i'j'\ldots}^{a'b'\ldots} \c{a'}\c{b'}\ldots\a{j'}\a{i'} , \quad t_{xy\ldots}^{wz\ldots} = 0
\end{align}
where the operator $T$ generates all internally-contracted excitations between core, active, and external orbitals. \cref{eq:mr_adc_wfn} is equivalent to the wavefunction used in internally-contracted multi-reference unitary coupled cluster theory and the related approaches.\cite{Kirtman:1981p798,Hoffmann:1988p993,Yanai:2006p194106,Yanai:2007p104107,Chen:2012p014108,Li:2015p2097,Li:2016p164114} As we will discuss in \cref{sec:comp_details}, excitations generated by the operator $T$ are linearly-dependent and the redundant amplitudes need to be eliminated.\cite{Yanai:2007p104107,Evangelista:2011p114102,Hanauer:2011p204111,Datta:2012p204107} A procedure for removing the redundant amplitudes is described in the Appendix.

To define the MR-ADC perturbative series, we must choose the zeroth-order Hamiltonian $H^{(0)}$. While the choice of $H^{(0)}$ is flexible, it must guarantee that the requirement 2 is satisfied. This suggests that $H^{(0)}$ must be an interacting Hamiltonian in the active space, rather than a Fock-like one-electron operator. In our MR-ADC development we choose $H^{(0)}$ to be the Dyall Hamiltonian,\cite{Dyall:1995p4909,Angeli:2001p10252,Angeli:2001p297,Angeli:2004p4043} defined as:
\begin{align}
	\label{eq:h_dyall_general}
	H^{(0)} \equiv C + \sum_{ij} \f{i}{j} \c{i}\a{j} + \sum_{ab} \f{a}{b} \c{a}\a{b} + H_{act}
\end{align}
where
\begin{align}
	\label{eq:f_gen}
	\f{p}{q} &= \h{p}{q} + \sum_{rs} \v{pr}{qs} \pdm{s}{r} \ , \quad \pdm{q}{p} = \braket{\Psi_0|\c{p}\a{q}|\Psi_0} \\
	\label{eq:h_act}
	H_{act} &= \sum_{xy}(\h{x}{y} + \sum_{i} \v{xi}{yi}) \c{x} \a{y} + \frac{1}{4} \sum_{xywz} \v{xy}{zw} \c{x} \c{y} \a{w} \a{z} \\
	C &= \sum_i \h{i}{i} + \frac{1}{2}\sum_{ij}\v{ij}{ij} - \sum_i  \f{i}{i} = E_{fc} - \sum_i  \f{i}{i} 
\end{align}
The Dyall Hamiltonian $H^{(0)}$ includes all active-space terms of the full electronic Hamiltonian $H$ (\cref{eq:hamiltonian}), satisfying the eigenvalue problem
\begin{align}
	\label{eq:h_zero_eval_problem}
	\braket{\Psi_J | H^{(0)} |\Psi_I} = \braket{\Psi_J | H | \Psi_I} = E_I \delta_{IJ}
\end{align}
for all CASCI/CASSCF model wavefunctions $\ket{\Psi_I}$ with eigenvalues $E_I$, which is a necessary condition for fulfilling the requirement 2. For convenience, we work in the basis of the diagonal core and external generalized Fock operators ($\f{i}{j} \rightarrow \e{i} \d{i}{j}$, $\f{a}{b} \rightarrow \e{a} \d{a}{b}$), where $H^{(0)}$ takes the form:
\begin{align}
	\label{eq:h_dyall_diagonal}
	H^{(0)} &= C + \sum_{i} \e{i} \c{i}\a{i} + \sum_{a} \e{a} \c{a} \a{a} + H_{act}
\end{align}

We now consider the expansion of the propagator with respect to the perturbation $V = H - H^{(0)}$ that defines the MR-ADC approximations. Conveniently, the general form of the MR-ADC equations is equivalent to that of SR-ADC presented in \cref{sec:theory_background_sr_adc}. The $n$-th order MR-ADC approximation, which will be termed as MR-ADC(n) henceforth, is defined by truncating the expansion for $\mathbf{G}_{+}(\omega)$ in \cref{eq:g_pt_series} after $\mathbf{G}_{+}^{(n)}(\omega)$. The $n$-th-order contribution $\mathbf{G}_{+}^{(n)}(\omega)$ is given by \cref{eq:Gn_matrix,eq:T_matrix,eq:A_matrix,eq:S_matrix,eq:M_matrix} and the MR-ADC(n) eigenvalue problem can be written as in \cref{eq:adc_eig_problem}. Despite general similarities, in MR-ADC, the matrices $\mathbf{T}^{(n)}_{+}$, $\mathbf{M}_{+}^{(n)}$, and $\mathbf{S}_{+}^{(n)}$ that compose $\mathbf{G}_{+}^{(n)}(\omega)$ are evaluated using the {\it multi-reference} model wavefunctions $\ket{\Psi_I}$, which incorporate the information about the active-space correlation and electronic states into the description of the propagator. 

\subsection{Perturbative analysis of the MR-ADC equations}
\label{sec:theory_mr_adc_pert_analysis}

In this section, we perform a perturbative analysis of the MR-ADC equations to illustrate the most important features of this theory. To compute the MR-ADC(n) excitation energies and transition amplitudes, we need to evaluate the $\mathbf{M}_{+}$, $\mathbf{S}_{+}$, and $\mathbf{T}_{+}$ matrices up to the $n$-th order in perturbation theory. This requires: (i) deriving expressions for the effective Hamiltonian $\tilde{H}^{(k)}$, (ii) constructing the operator manifolds $\mathbf{h}_{+}^{(k)\dag}$, (iii) solving equations for the $k$-th-order contributions to the $t_{i'j'\ldots}^{a'b'\ldots}$ amplitudes (\cref{eq:mr_adc_t_amplitudes}), and (iv) evaluating the transformed operators $\mathbf{\tilde{q}}^{(k)\dag}$. Since for the polarization propagator $\mathbf{G}_{+}(\omega)$ and $\mathbf{G}_{-}(\omega)$ contain the same information, we will only consider $\mathbf{G}_{+}(\omega)$ and drop the subscript $+$ everywhere in the equations.

\subsubsection{Zeroth-order contributions}
The zeroth-order operators $\tilde{H}^{(0)}$ and $\mathbf{\tilde{q}}^{(0)\dag}$ have a simple form:
\begin{align}
	\label{eq:H_bch_0}
	\tilde{H}^{(0)} &= H^{(0)} \\
	\label{eq:q_bch_0}
	\tilde{q}^{(0)\dag}_\mu &= q^\dag_\mu = a_{q}^{p} - \braket{\Psi_0|a_{q}^{p}|\Psi_0}
\end{align}
where we define $a_{q}^{p} \equiv \c{p}\a{q}$.
Following procedure outlined in \cref{sec:theory_background_sr_adc}, from \cref{eq:q_bch_0} we determine that the zeroth-order operators $\mathbf{h}^{(0)\dag}$ have the form of the single-particle operators $a_{q}^{p}$. Importantly, $\mathbf{h}^{(0)\dag}$ must satisfy VAC with respect to the ground-state model wavefunction $\ket{\Psi_0}$, i.e.\@ \mbox{$h^{(0)}_{\mu}\ket{\Psi_0} = 0$}. Since $\ket{\Psi_0}$ is a CAS-type wavefunction, there is a total of nine different classes of the operators $a_{q}^{p}$, where indices $p$ and $q$ belong to different orbitals subspaces (i.e., core, active, or external). Out of nine classes, two classes with only core ($a^{i}_{j}$) or only external ($a^{a}_{b}$) indices are redundant as they do not produce excited configurations when acting on $\ket{\Psi_0}$. We do not include these operators in the operator manifold $\mathbf{h}^{(0)\dag}$. Among the remaining types, the operators $a^{x}_{i}$, $a^{a}_{i}$, and $a^{a}_{x}$ can be added to $\mathbf{h}^{(0)\dag}$, while their adjoints contribute to $\mathbf{h}^{(0)}$. The last operator class with all active indices ($a^{x}_{y}$) generates excitations in the active space, but does not satisfy VAC with respect to $\ket{\Psi_0}$, and thus cannot be included in $\mathbf{h}^{(0)\dag}$. These operators require a special treatment in our development.

Expanding the active-space operators in the form $a^{x}_{y} = \sum_{I} Z^\dag_I c_{I,xy}$, where $Z^\dag_I$ is a complete set of the active-space eigenoperators\cite{Kutzelnigg:1998p5578} with a property $Z^\dag_I \ket{\Psi_0} = \ket{\Psi_I}$, we express the configurations generated by $a^{x}_{y}$ as:
\begin{align}
	\label{eq:xy_as_z_act_hilbert}
	a^{x}_{y}\ket{\Psi_0} = \sum_{I} Z^\dag_I \ket{\Psi_0} c_{I,xy} = \sum_{I} \ket{\Psi_I}\braket{\Psi_I|a^{x}_{y}|\Psi_0}
\end{align}
where the r.h.s.\@ of \cref{eq:xy_as_z_act_hilbert} is obtained by inserting the resolution of identity over a complete set of the active-space model states $\ket{\Psi_I}$ in the l.h.s.\@ of that equation. \cref{eq:xy_as_z_act_hilbert} suggests the form for the coefficients $c_{I,xy} = \braket{\Psi_I|a^{x}_{y}|\Psi_0}$ and the eigenoperators\cite{Lowdin:1985p285} $Z^\dag_I$:
\begin{align}
	\label{eq:z_ketbra}
	Z^\dag_I &= \ket{\Psi_I}\bra{\Psi_0} 
\end{align}
Since the CASCI/CASSCF model states $\ket{\Psi_I}$ are orthogonal, the excited-state eigenoperators $Z^\dag_I$ ($I > 0$) automatically fulfill VAC
\begin{align}
	\label{eq:z_act}
	Z_I \ket{\Psi_0} &= 0 \qquad (I > 0)
\end{align}
and thus can be included in the operator manifold $\mathbf{h}^{(0)\dag}$. Importantly, choosing $Z^\dag_I$ in the form of \cref{eq:z_ketbra} together with the choice for the zeroth-order Hamiltonian (\cref{eq:h_dyall_diagonal}) ensures that the requirement 2 discussed in \cref{sec:theory_mr_adc_general_aspects} is satisfied. The completeness of the operator set $Z^\dag_I$ is equivalent to the availability of the complete model space $\ket{\Psi_I}$. For small active spaces, it is possible to operate with a complete set of $\ket{\Psi_I}$ such that the set of $Z^\dag_I$ is complete. However, in most computations, it will be necessary to truncate the set $\ket{\Psi_I}$ to include only the low-energy states of interest, introducing an approximation (see \cref{sec:comp_details} for details).

We summarize that the $\mathbf{h}^{(0)\dag}$ operator manifold consists of the four classes of operators:
\begin{align}
\mathbf{h}^{(0)\dag} = \left\{Z^\dag_I; a^{a}_{i}; a^{x}_{i}; a^{a}_{x} \right\}    
\end{align}
The zeroth-order MR-ADC matrices have the general form:
\begin{align}
	\mathbf{M}^{(0)} & = \{\mathbf{h}^{(0)}|\tilde{\mathcal{H}}^{(0)}|\mathbf{h}^{(0)}\}  \\
	\mathbf{S}^{(0)} & = \{\mathbf{h}^{(0)}|\mathbf{h}^{(0)}\}  \\
	\mathbf{T}^{(0)} & = \{\mathbf{\tilde{q}}^{(0)}|\mathbf{h}^{(0)}\} 
\end{align}
Explicit equations for $\mathbf{M}^{(0)}$ and $\mathbf{S}^{(0)}$ are shown in the Supporting Information. In contrast to SR-ADC, these matrices are non-diagonal, due to the non-orthogonal nature of the internally-contracted configurations (e.g., $a^{x}_{i}\ket{\Psi_0}$) and the non-diagonal form of the Dyall Hamiltonian. However, the off-diagonal blocks of $\mathbf{M}^{(0)}$ and $\mathbf{S}^{(0)}$ corresponding to different types of $\mathbf{h}^{(0)\dag}$ vanish. 

\begin{figure*}[t!]
	\includegraphics[width=0.7\textwidth]{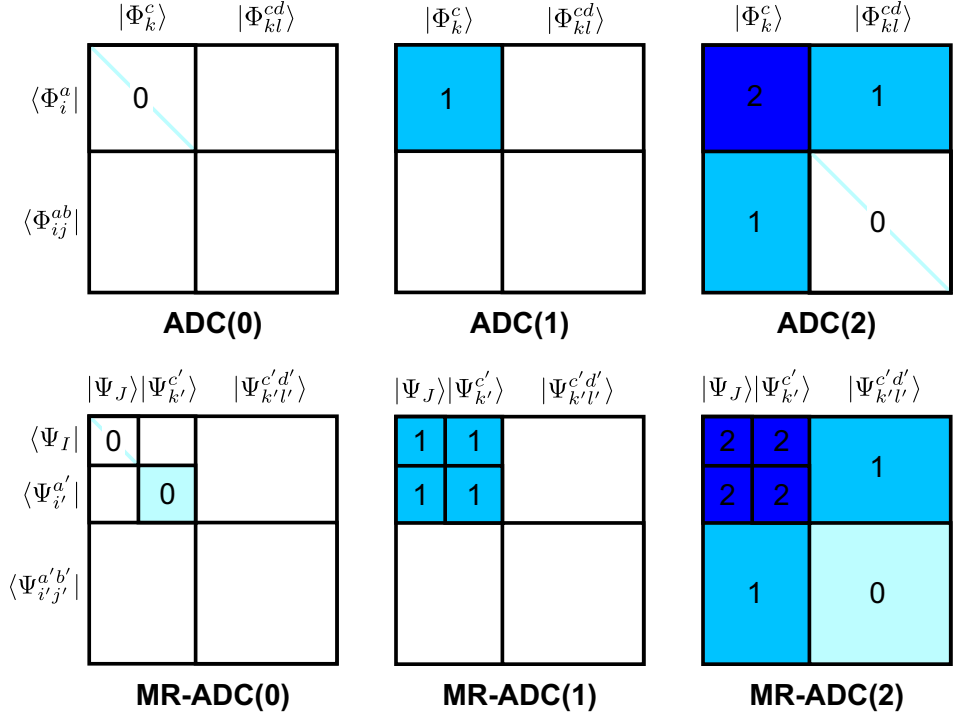}
	\captionsetup{justification=raggedright,singlelinecheck=false}
	\caption{Structure of the effective Liouvillean matrix $\mathbf{M}$ in the ADC(n) and MR-ADC(n) approximations (n = 0, 1, 2). Non-zero elements of $\mathbf{M}$ are highlighted in color, numbers indicate the perturbation order at which the expansion of effective Hamiltonian $\tilde{H}$ is truncated for each block. Wavefunctions $\ket{\Phi_i^a}$ and $\ket{\Phi_{ij}^{ab}}$ are singly and doubly excited determinants, $\ket{\Psi_I}$ are excited CASCI model states, $\ket{\Psi_{i'}^{a'}}$ and $\ket{\Psi_{i'j'}^{a'b'}}$ are singly and doubly excited multi-reference states with at least one core or external index.}
	\label{fig:adc_M}
\end{figure*}                                                                                                                                                                             

\subsubsection{First-order contributions}
Expanding the BCH expansions in \cref{eq:q_bch,eq:H_bch} to the first order, we obtain:
\begin{align}
	\label{eq:H_bch_1}
	\tilde{H}^{(1)} &= V + [H^{(0)}, T^{(1)} - T^{\dag(1)}]  \\
	\label{eq:q_bch_1}
	\tilde{q}^{(1)\dag}_\mu &= [a_{q}^{p}, T^{(1)} - T^{\dag(1)}]
\end{align}
Since $V$ is a two-electron operator, the first-order excitation operator $T^{(1)}$ only includes up to the two-body terms: $T^{(1)} = T_1^{(1)} + T_2^{(1)}$. From \cref{eq:q_bch_1} we determine the first-order operator manifold $\mathbf{h}^{(1)\dag}$, which consists of eight different types of double excitations:
\begin{align}
\mathbf{h}^{(1)\dag} = &\left\{ a_{ij}^{ab}; a_{ij}^{ax}; a_{ix}^{ab}; a_{ij}^{xy}; a_{xy}^{ab}; a_{ix}^{yz}; a_{xy}^{az}; a_{ix}^{ay} \right\} 
\end{align}
where $a_{rs}^{pq} \equiv \c{p}\c{q}\a{s}\a{r}$.
The MR-ADC matrices have the following structure:
\begin{align}
	\mathbf{M}^{(1)} & = \{\mathbf{h}^{(0)}|\tilde{\mathcal{H}}^{(1)}|\mathbf{h}^{(0)}\} + \{\mathbf{h}^{(1)}|\tilde{\mathcal{H}}^{(0)}|\mathbf{h}^{(0)}\} \notag \\
	&+ \{\mathbf{h}^{(0)}|\tilde{\mathcal{H}}^{(0)}|\mathbf{h}^{(1)}\} \\
	\mathbf{S}^{(1)} & = \{\mathbf{h}^{(1)}|\mathbf{h}^{(0)}\} + \{\mathbf{h}^{(0)}|\mathbf{h}^{(1)}\}  \\
	\mathbf{T}^{(1)} & = \{\mathbf{\tilde{q}}^{(1)}|\mathbf{h}^{(0)}\} + \{\mathbf{\tilde{q}}^{(0)}|\mathbf{h}^{(1)}\} 
\end{align}
In the MR-ADC(1) approximation, where $\mathbf{M} \approx \mathbf{M}^{(0)} + \mathbf{M}^{(1)}$, the terms that depend on the double-excitation manifold $\mathbf{h}^{(1)}$ do not contribute, since the $\{\mathbf{h}^{(1)}|\tilde{\mathcal{H}}^{(0)}|\mathbf{h}^{(1)}\}$ block of the effective Liouvillean matrix is a contribution to $\mathbf{M}^{(2)}$. However, these $\mathbf{h}^{(1)}$ contributions need to be included in MR-ADC(2) and the higher-order approximations. 
Evaluation of $\mathbf{M}^{(1)}$ requires the first-order single-excitation ($t_{i}^{a(1)}$, $t_{x}^{a(1)}$, $t_{i}^{x(1)}$) and (semi-internal) double-excitation ($t_{ix}^{ay(1)}$, $t_{xy}^{aw(1)}$, $t_{ix}^{yw(1)}$) amplitudes. These amplitudes can be obtained by solving a system of the projected amplitude equations 
\begin{align}
	\label{eq:proj_amplitude_equations_1}
	\braket{\Psi_0|a_{q}^{p}\tilde{H}^{(1)}|\Psi_0} = 0 \\
	\label{eq:proj_amplitude_equations_2}
	\braket{\Psi_0|a_{rs}^{pq}\tilde{H}^{(1)}|\Psi_0} = 0 
\end{align}
where $a_{q}^{p} \in \{ a_{a}^{i}; a_{a}^{x} ; a_{x}^{i}\}$ and $a_{rs}^{pq} \in \{ a_{ay}^{ix} ; a_{aw}^{xy} ; a_{yw}^{ix} \}$. We discuss the solution of these equations in more detail in the Appendix.

\subsubsection{Second-order contributions}

The second-order operators $\tilde{H}^{(2)}$ and $\mathbf{\tilde{q}}^{(2)\dag}$ have the form:
\begin{align}
	\label{eq:H_bch_2}
	\tilde{H}^{(2)} &= [H^{(0)}, T^{(2)} - T^{\dag(2)}] \notag \\
	&+ \frac{1}{2}[V + \tilde{H}^{(1)}, T^{(1)} - T^{\dag(1)}] \\
	\label{eq:q_bch_2}
	\tilde{q}^{(2)\dag}_\mu &= [a_{q}^{p}, T^{(2)} - T^{\dag(2)}] \notag \\
	&+ \frac{1}{2} [[a_{q}^{p}, T^{(1)} - T^{\dag(1)}], T^{(1)} - T^{\dag(1)}]
\end{align}
The second term in \cref{eq:q_bch_2} generates the three-body operators that compose the second-order operator manifold $\mathbf{h}^{(2)\dag}$. 
The MR-ADC matrices contain the following second-order terms:
\begin{align}
	\mathbf{M}^{(2)} & = \{\mathbf{h}^{(0)}|\tilde{\mathcal{H}}^{(2)}|\mathbf{h}^{(0)}\} + \{\mathbf{h}^{(1)}|\tilde{\mathcal{H}}^{(1)}|\mathbf{h}^{(0)}\} \notag \\
	& + \{\mathbf{h}^{(0)}|\tilde{\mathcal{H}}^{(1)}|\mathbf{h}^{(1)}\} + \{\mathbf{h}^{(1)}|\tilde{\mathcal{H}}^{(0)}|\mathbf{h}^{(1)}\} \notag \\
	& + \{\mathbf{h}^{(2)}|\tilde{\mathcal{H}}^{(0)}|\mathbf{h}^{(0)}\} + \{\mathbf{h}^{(0)}|\tilde{\mathcal{H}}^{(0)}|\mathbf{h}^{(2)}\} \\
	\mathbf{S}^{(2)} & =  \{\mathbf{h}^{(1)}|\mathbf{h}^{(1)}\} + \{\mathbf{h}^{(2)}|\mathbf{h}^{(0)}\} + \{\mathbf{h}^{(0)}|\mathbf{h}^{(2)}\} \\
	\mathbf{T}^{(2)} & =  \{\mathbf{\tilde{q}}^{(1)}|\mathbf{h}^{(1)}\} + \{\mathbf{\tilde{q}}^{(2)}|\mathbf{h}^{(0)}\} + \{\mathbf{\tilde{q}}^{(0)}|\mathbf{h}^{(2)}\} 
\end{align}
Terms containing $\mathbf{h}^{(2)}$ need to be included only in MR-ADC(4) and higher-order approximations. However, all terms involving the double-excitation manifold $\mathbf{h}^{(1)}$ must be included already in MR-ADC(2). Computation of $\mathbf{M}^{(2)}$ requires solving for all amplitudes of the $T_1^{(1)}$ and $T_2^{(1)}$ operators, as well as for the single- and semi-internal double-excitation amplitudes of $T_1^{(2)}$ and $T_2^{(2)}$.

\cref{fig:adc_M} compares the effective Liouvillean matrix $\mathbf{M}$ for the single-reference ADC(n) and multi-reference MR-ADC(n) approximations (n = 0, 1, 2). Although in MR-ADC there are more excitation classes than in SR-ADC, the perturbative structure of the MR-ADC(n) and ADC(n) matrices is very similar. These matrices become equivalent in the limit of the single-determinant $\ket{\Psi_{0}}$ and zero active orbitals. 

Finally, we comment about size-extensivity of the MR-ADC(n) approximations. The commutator structure of the MR-ADC equations ensures that all of the terms that appear in these equations are fully linked, which formally guarantees size-consistency of the MR-ADC excitation energies and transition amplitudes at any level of approximation, similar to that of the single-reference ADC. In practice, however, size-consistency can be violated during the numerical solution of the MR-ADC equations as a result of removing redundant excitations of the operator $T$. These size-consistency errors originate from the disconnected terms involving the single and semi-internal amplitudes (e.g., $t_{i}^{a(1)}$ and $t_{ix}^{ay(1)}$) that become non-zero when linear dependencies are eliminated.\cite{Hanauer:2011p204111,Hanauer:2012p131103} We will investigate the size-consistency errors of the MR-ADC(1) approximation in \cref{sec:size_consistency}.

\section{Implementation: MR-ADC(1)}
\label{sec:implementation}

We have derived and implemented equations for the MR-ADC(1) approximation as a first step in constructing the hierarchy of the MR-ADC methods. The general structure of the MR-ADC(1) approximation was described in \cref{sec:theory_mr_adc_pert_analysis} and the explicit equations are shown in the Supporting Information. Although the single-reference ADC(1) energies are equivalent to those obtained from the Tamm-Dancoff approximation\cite{Fetter2003} (TDA), the MR-ADC(1) and the multi-configurational TDA (MC-TDA)\cite{Yeager:1979p77,Dalgaard:1980p816,Radojevic:1985p2991,Sangfelt:1998p4523,Nakatani:2014p024108}  energies are in general different. This is because in MR-ADC(1) the effective Liouvillean matrix contains additional non-zero terms that depend on the single-excitation ($t_{x}^{a(1)}$, $t_{i}^{x(1)}$) and semi-internal ($t_{ix}^{ay(1)}$, $t_{xy}^{aw(1)}$, $t_{ix}^{yw(1)}$) amplitudes that are not present in MC-TDA. Neglecting these terms reduces the MR-ADC(1) equations to those of MC-TDA. 

The main steps of the MR-ADC(1) implementation for a CASSCF reference wavefunction are summarized below:
\begin{enumerate}
\item Choose active space, compute the ground-state CASSCF wavefunction $\ket{\Psi_0}$.
\item Using the optimized CASSCF orbitals, compute the energies $E_I$ and the wavefunctions $\ket{\Psi_I}$ for $N_{\mathrm{CAS}}$ lowest-energy CASCI states. 
\item Compute active-space reduced density matrices (RDMs) for the ground state $\ket{\Psi_0}$ and transition RDMs between $\ket{\Psi_0}$ and the excited CASCI states $\ket{\Psi_I}$ ($I > 0$).
\item Solve linear equations for the single-excitation ($t_{x}^{a(1)}$, $t_{i}^{x(1)}$) and semi-internal ($t_{ix}^{ay(1)}$, $t_{xy}^{aw(1)}$, $t_{ix}^{yw(1)}$) amplitudes (see the Appendix for details).
\item Compute the overlap ($\mathbf{S} = \mathbf{S}^{(0)}$) and the effective Liouvillean matrices ($\mathbf{M} = \mathbf{M}^{(0)} + \mathbf{M}^{(1)}$).
\item Solve the generalized eigenvalue problem \eqref{eq:adc_eig_problem} to compute excitation energies. 
\end{enumerate}
In the algorithm outlined above, the model space should contain all CASCI states $\ket{\Psi_I}$ that are important for the problem and spectral region of interest. Evaluation of the $\mathbf{M}$ matrix elements requires computation of up to the three-particle ground-state RDM (3-RDM) and three-particle transition RDM, which have $\mathcal{O}(N_{\mathrm{CAS}} \times N_{\mathrm{det}} \times N^6_{\mathrm{act}})$ computational scaling, where $N_{\mathrm{det}}$ is the dimension of the active-space Hilbert space and $N_{\mathrm{act}}$ is the number of active orbitals. In addition, solving the amplitude equations for the semi-internal excitations with three active-space indices ($t_{xy}^{aw(1)}$ and $t_{ix}^{yw(1)}$) requires computing the ground-state 4-RDM, which has $\mathcal{O}(N_{\mathrm{det}} \times N^8_{\mathrm{act}})$ computational cost. In our implementation, we avoid computation and storage of 4-RDM using the imaginary-time propagation algorithm outlined in the Appendix.

\section{Computational details}
\label{sec:comp_details}
We implemented MR-ADC(1) in a standalone Python program. To obtain one- and two-electron integrals and the CASSCF reference wavefunctions, our program was interfaced with \textsc{Pyscf}.\cite{Sun:2018pe1340} The main steps of our implementation are described in \cref{sec:implementation}. In all MR-ADC(1) computations, the CASSCF reference molecular orbitals were optimized for the ground electronic state with tight convergence parameters for the energy (10$^{-8}$ \eh). These orbitals were used in the following CASCI computation, which produced wavefunctions for the excited states in the active space. We refer to this procedure as CASCI/CASSCF. We denote active spaces used in CASCI/CASSCF as ($n$e, $m$o), where $n$ is the number of active electrons and $m$ is the number of orbitals. The MR-ADC(1) results were benchmarked against accurate reference data from full configuration interaction (FCI) and density matrix renormalization group (DMRG) and were compared to results from strongly-contracted $N$-electron valence second-order perturbation theory (sc-NEVPT2)\cite{Angeli:2001p10252} and its quasidegenerate variant (sc-QD-NEVPT2).\cite{Angeli:2004p4043} The NEVPT2 computations were performed using the \textsc{Orca} program\cite{Neese:2017pe1327} and employed the state-averaged CASSCF orbitals (SA-CASSCF). 

In addition to choosing the active space, the MR-ADC(1) results depend on three parameters: (i) the parameter $\Delta_{conv}$ for the imaginary-time propagation used to compute the semi-internal amplitudes of the effective Hamiltonian, (ii) the thresholds $\eta^{[0']}$ and $\eta^{[\pm1']}$ for removing linear dependencies in the overlap matrices, and (iii) $N_{\mathrm{CAS}}$, the number of the CASCI states in the model space. 

We refer the interested readers to the Appendix for details about the imaginary-time propagation and the overlap matrices. In short, our imaginary-time algorithm follows the procedure described in Ref.\@ \citenum{Sokolov:2016p064102}. The time propagation is performed using the embedded Runge-Kutta algorithm, which automatically determines the time step based on the accuracy parameter $\Delta_{conv}$.\cite{Press:2007} Since the imaginary-time propagation is a relatively inexpensive step of our algorithm, we use a small value $\Delta_{conv}=10^{-6}$ in all computations, which allows to compute numerically accurate semi-internal amplitudes with $\sim$ 10 to 40 imaginary-time steps. 

To remove linear dependencies, we first diagonalize the overlap matrices $\mathbf{S^{[0']}}$, $\mathbf{S^{[+1']}}$, and $\mathbf{S^{[-1']}}$ defined in \cref{eq:S_0p,eq:S_-1p,eq:S_+1p}, respectively. We then arrange the resulting eigenvalues $s^{[i]}_{p}$ ($i$ = $0'$, $+1'$, $-1'$) in the ascending order and project out the eigenvectors with the smallest $s^{[i]}_{p}$ that satisfy the following condition:
\begin{align}
\frac{\sum_{p}^{trunc}s^{[i]}_{p}}{\sum_{p}s^{[i]}_{p}} \le \eta^{[i]}
\end{align}
where the sum in the numerator runs over the truncated $s^{[i]}_{p}$ and the denominator contains the total sum of $s^{[i]}_{p}$. In practice, the diagonalization needs to be performed only for the active-space blocks of $\mathbf{S^{[0']}}$, $\mathbf{S^{[+1']}}$, and $\mathbf{S^{[-1']}}$ with the same core ($i=j$) and external ($a=b$) indices. Our numerical tests indicate that numerical instabilities due to the linear dependencies can be completely eliminated when $\eta^{[0']}$ and $\eta^{[\pm1']}$ are chosen to be 10$^{-8}$ and 10$^{-3}$, respectively, which is consistent with the values used in implementations of other internally-contracted multi-reference theories.\cite{Yanai:2007p104107,Evangelista:2011p114102,Hanauer:2011p204111,Datta:2012p204107} We use these values in all of our computations. We note that the procedure for removing linear dependencies used in this work treats redundancies in single and double (semi-internal) excitations on equal footing, but is not unique. Other strategies for eliminating linear dependencies can be used, where either one of the excitation classes is omitted or excitations of one class are removed from excitations of another class.\cite{Hanauer:2011p204111} We refer to the work by Hanauer and K\"ohn in Ref.\@ \citenum{Hanauer:2011p204111} for a detailed numerical analysis of these alternative methods within the framework of internally-contracted multi-reference coupled cluster theory.

\begin{table*}[t!]
	\captionsetup{justification=raggedright,singlelinecheck=false}
	\caption{Size-consistency errors of the MR-ADC(1) excitation energies (eV) for a system consisting of two identical water molecules separated from each other by 10000 $\angstrom$ (cc-pVDZ basis set). The errors are computed as: $\Delta E = E(2$H$_2$O$) - 2E($H$_2$O$)$, where $E(2$H$_2$O$)$ and $E($H$_2$O$)$ are the dimer and monomer excitation energies for each state. Results are shown for the fixed H--O--H angle (104.5\degree), two different O--H bond lengths ($r$ = 1.0 and 2.0 \angstrom), different values of the overlap truncation parameter ($\eta^{[\pm1']}$, see \cref{sec:comp_details} for details), and four lowest-energy singlet states (S$_n$). The \mbox{(8e, 8o)} and \mbox{(4e, 4o)} active spaces were used for the dimer and monomer CASSCF reference wavefunctions, respectively. }
	\label{tab:size_consistency}
	\setstretch{1.1}
    \begin{tabular}{L{1cm}C{2.75cm}C{2.75cm}C{2.75cm}C{2.75cm}C{2.75cm}}
        \hline
        \hline
        	&  \multicolumn{5}{c}{Overlap truncation threshold ($\eta^{[\pm1']}$)}\\
        State & $10^{-1}$ & $10^{-2}$ & $10^{-3}$ & $10^{-4}$ & $10^{-5}$\\
        \hline
\multicolumn{6}{c}{$r = 1.0$ \angstrom} \\
        \hline       
S$_1$ & 4.7 $\times$ $10^{-3}$ & 8.0 $\times$ $10^{-3}$ & $-$9.7 $\times$ $10^{-4}$ & 2.8 $\times$ $10^{-2}$   & 1.1 $\times$ $10^{-2}$  \\
S$_2$ & 8.7 $\times$ $10^{-3}$ & 1.5 $\times$ $10^{-3}$ & $-$6.2 $\times$ $10^{-3}$ & 4.2 $\times$ $10^{-2}$   & $-$1.3 $\times$ $10^{-2}$  \\
S$_3$ & 4.3 $\times$ $10^{-3}$ & 1.1 $\times$ $10^{-2}$ & 2.0 $\times$ $10^{-3}$ &  8.7 $\times$ $10^{-3}$  &  1.8 $\times$ $10^{-2}$  \\
S$_4$ & 5.0 $\times$ $10^{-3}$ & 5.3 $\times$ $10^{-2}$ & $-$1.3 $\times$ $10^{-2}$ & $-$2.8 $\times$ $10^{-3}$   &$-$4.0 $\times$ $10^{-3}$  \\
        \hline       
\multicolumn{6}{c}{$r = 2.0$ \angstrom} \\
        \hline       
S$_1$ & 4.0 $\times$ $10^{-3}$ & 4.1 $\times$ $10^{-3}$  & 2.1 $\times$ $10^{-3} $&  2.2 $\times$ $10^{-3} $ & 2.2 $\times$ $10^{-3}  $\\
S$_2$ & 5.3 $\times$ $10^{-3}$ & 1.5 $\times$ $10^{-3}$  & $-$6.9 $\times$ $10^{-5}$ & $-$4.2 $\times$ $10^{-5}$ & $-$5.8 $\times$ $10^{-5}$  \\
S$_3$ & 2.2 $\times$ $10^{-2}$ & 5.5 $\times$ $10^{-3}$  & 3.2 $\times$ $10^{-3} $&  3.3 $\times$ $10^{-3} $ & 3.2 $\times$ $10^{-3}  $\\
S$_4$ & 1.0 $\times$ $10^{-2}$ & 4.0 $\times$ $10^{-3}$  & 2.9 $\times$ $10^{-3} $&  2.9 $\times$ $10^{-3} $ & 2.8 $\times$ $10^{-3}  $\\		
        \hline
        \hline
    \end{tabular}
\end{table*}

Finally, the MR-ADC(1) results depend on the number of the CASCI states included in the model space ($N_{\mathrm{CAS}}$, see \cref{sec:implementation}). The optimal value of $N_{\mathrm{CAS}}$ depends on the system and should include all active-space states in the energy range of interest. In our computations, we usually start with $N_{\mathrm{CAS}}=10$ and increase it until the excitation energies for the relevant states are converged. For the systems and active spaces considered in this work, the optimal value of $N_{\mathrm{CAS}}$ ranged from 20 to 50 states, with the exception of the Be atom with the $(2s3s2p3p3d)$ active space, where we had to use $N_{\mathrm{CAS}}$ = 80 to obtain converged energies for all excited states. An important feature of MR-ADC(1) is that the excited-state CASCI wavefunctions are only used to compute the transition reduced density matrices and thus can be discarded after their computation is complete.

\section{Results}
\label{sec:results}

\subsection{Size-consistency of excitation energies}
\label{sec:size_consistency}

We first examine size-consistency of the MR-ADC(1) energies for a system of two non-interacting water molecules. As we discussed in \cref{sec:theory_mr_adc_pert_analysis}, the MR-ADC approximations are formally size-consistent, but removing redundancies from the cluster operator $T$ can give rise to size-consistency errors. \cref{tab:size_consistency} shows size-consistency errors ($\Delta E$ in eV) of the MR-ADC(1) excitation energies for the four lowest-energy singlet excited states of the non-interacting water molecules with two different O--H bond lengths (1.0 and 2.0 \angstrom). Our numerical tests indicate that the size-consistency errors originate from the terms that depend on the core-active and active-external amplitudes (denoted as $[+1']$ and $[-1']$ in the Appendix) and vanish when these amplitudes are set to zero. To study the effect of the overlap truncation on the magnitude of the size-consistency errors, we computed $\Delta E$ for different values of the overlap truncation parameter $\eta^{[\pm1']}$ (\cref{sec:comp_details}). For all values of $\eta^{[\pm1']}$, the $\Delta E$ errors are small ($\sim$ $10^{-3}$ eV) with the largest error of only 0.053 eV. Changing $\eta^{[\pm1']}$ from $10^{-1}$ to $10^{-5}$ does not significantly affect the $\Delta E$ values, most of them change by less than an order of magnitude. 

We note that the observed size-consistency errors are intrinsic to the procedure used to project out linear dependencies and are not unique to MR-ADC. For example, Hanauer and K\"ohn\cite{Hanauer:2011p204111} demonstrated that eliminating redundancies in the equations of internally-contracted multi-reference coupled cluster theory also gives rise to size-extensivity and size-consistency errors. They developed alternative truncation schemes that allow to reduce or eliminate size-extensivity errors.\cite{Hanauer:2012p131103} These schemes can be readily adopted in MR-ADC to remove size-consistency errors, which will be the subject of our future work.

\subsection{Excitation energies of the \ce{Be} atom}

\begin{table*}[t!]
	\captionsetup{justification=raggedright,singlelinecheck=false}
	\caption{Excitation energies (eV) of the Be atom computed using CASCI and sc-NEVPT2 employing the $(2s3s2p3p3d)$ active space. The basis set and full configuration results (FCI) are from Ref.\@ \citenum{Graham:1998p6544}. The CASCI results were computed using the ground-state CASSCF orbitals, as well as the CASSCF orbitals averaged over $n$ lowest-energy states [SA($n$)-CASSCF, $n$ = 9 and 20]. States included in state-averaging are indicated with asterisks. Also shown are mean absolute errors (\mae) and standard deviations (\std) of the results, relative to FCI.}
	\label{tab:be_casci_nevpt}
	\setstretch{1}
	\small
    \begin{tabular}{L{2cm}C{2cm}C{3.25cm}C{3.25cm}C{3.25cm}C{2cm}}
        \hline
        \hline
        			& CASCI/ 		& CASCI/ 			& CASCI/ 			& sc-NEVPT2/ 		& FCI \\
        State 	& CASSCF 	& SA(9)-CASSCF 	& SA(20)-CASSCF 	& SA(20)-CASSCF 	& \\
        \hline
$2s^1 2p^1$ $^1P^0$  		& 6.08 	& 5.33$^*$	& 5.38$^*$	& 5.37$^*$  	&5.32  \\
$2s^1 3s^1$ $^1S$          		& 10.76 	& 6.75$^*$	& 6.69$^*$	& 6.77$^*$  	&6.77  \\
$2p^2$ $^1D$     			& 8.11 	& 7.29		& 7.04$^*$	& 7.14$^*$  	&7.09  \\
$2s^1 3p^1$ $^1P^0$  		& 17.21 	& 8.81		& 7.75		& 7.53  		&7.46  \\
$2s^1 3d^1$ $^1D$          		& 18.90 	& 12.76		& 11.16 		& 9.20  		&8.03  \\
$2s^1 2p^1$  $^3P^0$          	& 2.92 	& 2.69$^*$	& 2.63$^*$	& 2.73$^*$  	&2.73  \\
$2s^1 3s^1$  $^3S$          	& 11.94 	& 6.41$^*$	& 6.33$^*$	& 6.46$^*$  	&6.44  \\
$2s^1 3p^1$  $^3P^0$          	& 7.93 	& 7.42		& 7.24$^*$	& 7.33$^*$  	&7.30  \\
$2p^2$ $^3P$     		     	& 16.73 	& 8.38		& 7.39$^*$	& 7.45$^*$  	&7.42  \\
$2s^1 3d^1$  $^3D$           	& 16.47 	& 11.54		& 9.44		& 8.30  		&7.74  \\
\mae          				& 5.07 	& 1.13		& 0.56 		& 0.20   		& \\	
\std     	     				& 4.30 	& 1.74		& 1.08		& 0.38   		& \\	
        \hline
        \hline
    \end{tabular}
\end{table*}

\begin{table*}[t!]
	\captionsetup{justification=raggedright,singlelinecheck=false}
	\caption{Excitation energies (eV) of the Be atom computed using MR-ADC(1) with three active spaces employed in the reference CASSCF computation: $(2s2p)$, $(2s2p3s3p$), $(2s3s2p3p3d)$. The basis set and full configuration results (FCI) are from Ref.\@ \citenum{Graham:1998p6544}. Also shown are mean absolute errors (\mae) and standard deviations (\std) of the results, relative to FCI.}
	\label{tab:be}
	\setstretch{1}
	\small
\begin{threeparttable}
    \begin{tabular}{L{2cm}C{2.75cm}C{2.75cm}C{2.75cm}C{2cm}C{2.5cm}}
        \hline
        \hline
        		& MR-ADC(1) & MR-ADC(1) & MR-ADC(1) & FCI & Experiment\tnote{a}\\
        State & ($2s2p$) & ($2s2p3s3p$) & ($2s3s2p3p3d$) & & \\
        \hline
$2s^1 2p^1$ $^1P^0$  	& 5.43 & 5.31 & 5.37 & 5.32   & 5.28\\
$2s^1 3s^1$ $^1S$          	& 6.96 & 6.77 & 6.78 & 6.77   & 6.78\\
$2p^2$ $^1D$     		& 7.18 & 7.15 & 7.15 & 7.09   & 7.05\\
$2s^1 3p^1$ $^1P^0$  	& 7.62 & 7.46 & 7.48 & 7.46   & 7.46\\
$2s^1 3d^1$ $^1D$          	& 8.24  &8.07  &8.07  &8.03    &7.99 \\
$2s^1 4s^1$ $^1S$          	& 8.26 & 8.08 & 8.09 & 8.08   & 8.09\\
$2s^1 4p^1$ $^1P^0$  	& 8.48 & 8.30 & 8.32 & 8.30   & 8.31\\
$2s^1 4d^1$ $^1D$         	& 8.74  &8.55  &8.55  &8.54    &8.53 \\
$2s^1 5s^1$ $^1S$         	& 8.79 & 8.60 & 8.61 & 8.60   & 8.60\\
$2s^1 5p^1$ $^1P^0$  	& 8.88 & 8.69 & 8.70 & 8.69   & 8.69\\
$2s^1 6s^1$ $^1S$          	& 9.17 & 8.98 & 8.99 & 8.98   & 8.84\\
$2s^1 2p^1$  $^3P^0$          	&2.72  & 2.72 & 2.73 & 2.73   & 2.73\\
$2s^1 3s^1$  $^3S$          	&6.52  & 6.43 & 6.44 & 6.44   & 6.46\\
$2s^1 3p^1$  $^3P^0$          	&7.45  & 7.29 & 7.30 & 7.30   & 7.30\\
$2p^2$ $^3P$     		     	&7.65  & 7.65 & 7.55 & 7.42   & 7.40\\
$2s^1 3d^1$  $^3D$           	&7.91  & 7.74 & 7.75 & 7.74   & 7.69\\
$2s^1 4s^1$  $^3S$          	&8.15  & 7.98 & 7.99 & 7.99   & 8.00\\
$2s^1 4p^1$  $^3P^0$          	&8.45  & 8.27 & 8.28 & 8.27   & 8.28\\
$2s^1 4d^1$  $^3D$          	&8.63  & 8.45 & 8.46 & 8.45   & 8.42\\
$2s^1 5s^1$  $^3S$          	&8.74  & 8.56 & 8.57 & 8.56   & 8.56\\
$2s^1 5p^1$  $^3P^0$          	&8.87  & 8.68 & 8.69 & 8.69   & 8.69\\
$2s^1 6s^1$  $^3S$          	&9.06  & 8.88 & 8.90 & 8.89   & 8.82\\
$2s^1 6p^1$  $^3P^0$          	&9.14  & 8.96 & 8.97 & 8.96   & 8.89\\	
\mae          				&0.16  & 0.02 & 0.02 &    & \\	
\std     	     				&0.05  & 0.05 & 0.03 &    & \\	
        \hline
        \hline
    \end{tabular}
    \begin{tablenotes}
    \item[a] See Ref.\@ \citenum{Graham:1998p6544} for references to experimental results.
    \end{tablenotes}    
\end{threeparttable}
\end{table*}

In this section, we study the dependence of the MR-ADC(1) results on the size of the active space. In our benchmark, we consider the beryllium atom (Be), for which the accurate results from full configuration interaction (FCI) are available in the literature.\cite{Graham:1998p6544} We use the same basis set as in Ref.\@ \citenum{Graham:1998p6544} and employ three active spaces in our reference CASSCF computations, including all orbitals in parenthesis: $(2s2p)$, $(2s2p3s3p$), and $(2s3s2p3p3d)$. The largest active space corresponds to (2e, 13o).

We first investigate accuracy of the Be excitation energies computed using conventional methods, such as CASCI and strongly-contracted NEVPT2 (sc-NEVPT2). \cref{tab:be_casci_nevpt} shows results of CASCI and sc-NEVPT2 obtained using the largest $(2s3s2p3p3d)$ active space. The CASCI energies were computed using the ground-state CASSCF orbitals (CASCI/CASSCF), as well as the CASSCF orbitals averaged over $n$ lowest-energy states [CASCI/SA($n$)-CASSCF, $n$ = 9 and 20]. For sc-NEVPT2, only the SA($20$)-CASSCF orbitals were used. For all electronic states, the CASCI excitation energies strongly depend on the choice of molecular orbitals. In particular, using the ground-state CASSCF orbitals leads to very large mean absolute errors (\mae) and standard deviations (\std) of 5.07 and 4.30 eV, respectively, relative to FCI. These errors are substantially reduced when using the state-averaged CASSCF orbitals. In this case, the best agreement with the FCI benchmark results is observed for states that were included in state-averaging (errors of $\le$ 0.11 eV), while much larger errors (up to $\sim$ 3 eV) are obtained for the remaining states. Incorporating the description of dynamic correlation using the sc-NEVPT2/SA($20$)-CASSCF method further reduces the errors, yielding excitation energies in excellent agreement with FCI for states included in state-averaging. Although sc-NEVPT2 improves the description of the remaining states, their computed transition energies are significantly overestimated (up to 1.17 eV).

We now turn our attention to the MR-ADC(1) results presented in \cref{tab:be}. Importantly, while CASCI and sc-NEVPT2 can only be used to compute energies of transitions between orbitals in the active space, MR-ADC(1) provides information about excitations between all orbitals. Although MR-ADC(1) is the simplest approximation in the MR-ADC hierarchy, its accuracy can be improved by increasing the size of the active space. \cref{tab:be} demonstrates that the MR-ADC(1) excitation energies converge towards the FCI limit as the active space is expanded from $(2s2p)$ to $(2s3s2p3p3d)$. Including more orbitals in the active space also improves the description of excitations between active and external orbitals. For example, the error in excitation energy for the $2s^1 4p^1$  $^3P^0$ state reduces from 0.17 eV to 0.01 eV by increasing the active space from $(2s2p)$ to $(2s3s2p3p3d)$. Overall, for the largest $(2s3s2p3p3d)$ active space, the MR-ADC(1) results are in a close agreement with FCI, with mean absolute errors (\mae) and standard deviations (\std) of 0.02 and 0.03 eV, respectively. Although the MR-ADC(1) method used the ground-state CASSCF orbitals, its \mae and \std computed using the smallest ($2s2p$) active space (0.16 and 0.05 eV, respectively) are already smaller than those of sc-NEVPT2/SA($20$)-CASSCF employing the largest $(2s3s2p3p3d)$ active space (0.20 and 0.38 eV).

Interestingly, the MR-ADC(1) excitation energies agree very closely to the energies from multi-configurational linear response theory (MC-LR) reported in Ref.\@ \citenum{Graham:1998p6544} for almost all states but the $2s^1 2p^1$ $^1P^0$ state, where the difference of $\sim$ 0.05 eV is observed. Although both theories are based on the first-order approximation to the polarization propagator, MC-LR is formulated as a non-Hermitian eigenvalue problem, whereas the MR-ADC(1) energies are computed by diagonalizing a Hermitian matrix of a smaller dimension.

\subsection{Avoided crossing in \ce{LiF}}

Next, we test MR-ADC(1) for the description of an avoided crossing between the ground $X\, ^1\Sigma_g^+$ and the excited $2\, ^1\Sigma_g^+$ states of LiF.\cite{Finley:1998p299,Nakano:2002p1166,Angeli:2004p4043,Granovsky:2011p214113} At short bond distances, the wavefunctions of these two states are dominated by the ionic and covalent configurations, respectively. As the bond distance increases, the potential energy curves of the two states closely approach each other and their wavefunctions strongly interact, exchanging their character. In this section, we use the 6-31+G* basis set\cite{Hariharan:1973p213,Clark:1983p294} and compare the MR-ADC(1) results to those obtained from FCI, state-averaged CASSCF (SA-CASSCF), strongly-contracted quasidegenerate NEVPT2 (sc-QD-NEVPT2), and multi-configurational Tamm-Dancoff approximation (MC-TDA). As described in \cref{sec:implementation}, MC-TDA can be considered as an approximation to MR-ADC(1) where all terms that depend on the single- and semi-internal double-excitation amplitudes are neglected. The FCI results were computed using the semistochastic heat-bath configuration interaction algorithm (SHCI) implemented in the \textsc{Dice} program.\cite{Holmes:2016p3674,Sharma:2017p1595,Holmes:2017p164111} The $1s$ orbital of fluorine was not correlated in the SHCI computations. All active-space methods used the \mbox{(6e, 6o)} active space. In SA-CASSCF and sc-QD-NEVPT2, both $X\, ^1\Sigma_g^+$ and $2\, ^1\Sigma_g^+$ states were included in state-averaging to compute the reference CASSCF orbitals. In MR-ADC(1) and MC-TDA, the CASSCF orbitals were optimized for the ground $X\, ^1\Sigma_g^+$ state.

\begin{figure}[t!]
	\includegraphics[width=0.42\textwidth]{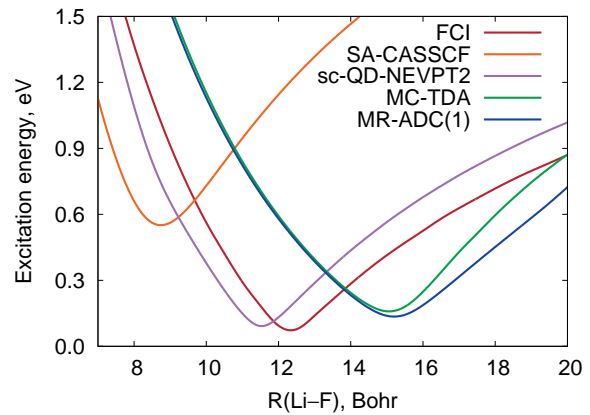}
	\captionsetup{justification=raggedright,singlelinecheck=false}
	\caption{Excitation energy (eV) of LiF between the $X\, ^1\Sigma_g^+$ and $2\, ^1\Sigma_g^+$ states as a function of the bond distance computed using FCI, SA-CASSCF, sc-QD-NEVPT2, MC-TDA, and MR-ADC(1) (6-31+G* basis set). The FCI energies were computed using the SHCI algorithm,\cite{Holmes:2016p3674,Sharma:2017p1595,Holmes:2017p164111} with the $1s$ orbital of fluorine atom not included in the correlation treatment. All active-space methods used the \mbox{(6e, 6o)} active space. In SA-CASSCF and sc-QD-NEVPT2, both states were included in state-averaging to compute the reference CASSCF orbitals. In MC-TDA and MR-ADC(1), the CASSCF orbitals were optimized for the $X\, ^1\Sigma_g^+$ state.}
	\label{fig:LiF}
\end{figure}          

\begin{table*}[t!]
	\captionsetup{justification=raggedright,singlelinecheck=false}
	\caption{Vertical excitation energies (eV) for the low-lying electronic states of \ce{C2} (\mbox{$r$ = 2.4 $a_0$}) computed using TDA, CASCI/CASSCF, MR-ADC(1), sc-NEVPT2, and DMRG (cc-pVDZ basis set). CASCI/CASSCF corresponds to a CASCI computation using the CASSCF ground-state orbitals. In CASCI/CASSCF, MR-ADC(1), and sc-NEVPT2, the CASSCF reference wavefunction used the \mbox{(8e, 8o)} active space. Also shown are mean absolute errors (\mae) and standard deviations (\std) of the results, relative to DMRG.}
	\label{tab:c2}
	\setstretch{1.1}
\begin{threeparttable}
    \begin{tabular}{L{1.5cm}C{2.5cm}C{3.5cm}C{2.5cm}C{2.5cm}C{2.5cm}}
        \hline
        \hline
        State & TDA & CASCI/CASSCF & MR-ADC(1) & sc-NEVPT2 & DMRG\tnote{a} \\
        \hline
        \(A\, {}^1\Pi_u\)               		& $-$1.30	& 2.31 & 1.97 & 1.42 & 1.29  \\
        \(B\, {}^1\Delta_g\)         		& \tnote{b}	& 4.08 & 4.08 & 2.45 & 2.17  \\
        \(B'\, {}^1\Sigma_g^+\)   		& \tnote{b}	& 4.07 & 4.04 & 2.73 & 2.45  \\
        \(C\, {}^1\Pi_g\)              		& 6.42	& 6.27 & 5.98 & 4.83 & 4.61  \\
        \(a\, {}^3\Pi_u\)              		& $-$2.41	& 0.97 & 0.64 & 0.34 & 0.21 \\
        \(b\, {}^3\Sigma_g^-\)         	& \tnote{b}	& 3.17 & 3.17 & 1.56 & 1.29  \\
        \(c\, {}^3\Sigma_u^+\)   		& $-$1.08	& 1.39 & 1.28 & 1.35 & 1.29  \\
        \(d\, {}^3\Pi_g\)				& \tnote{b}	& 3.74 & 3.57 & 2.85 & 2.65  \\
        \(1\, {}^3\Delta_u\)			& 7.35	& 8.03 & 7.17 & 7.02 & 6.66  \\
        \mae						& 2.01	& 1.27 & 1.03 & 0.21 &   \\
        \std						& 2.11	& 0.59 & 0.69 & 0.10 &   \\
        \hline
        \hline
    \end{tabular}
    \begin{tablenotes}
    \item[a] The DMRG results employing the \mbox{(12e, 28o)} active space are from Ref.\@ \citenum{Wouters:2014p1501}. 
    \item[b] Excited state with double excitation character, absent in TDA.
    \end{tablenotes}    
\end{threeparttable}
\end{table*}                                                                                                                                                                   

\cref{fig:LiF} plots the $X\, ^1\Sigma_g^+ \rightarrow 2\, ^1\Sigma_g^+$ excitation energy ($\Delta E$) for various bond distances computed using MR-ADC(1), MC-TDA, sc-QD-NEVPT2, SA-CASSCF, and FCI. The FCI curve exhibits a minimum corresponding to an avoided crossing at 12.25 $a_0$ with $\Delta E$ = 0.07 eV. SA-CASSCF shows the worst agreement with FCI predicting an avoided crossing at 8.75 $a_0$ with $\Delta E$ = 0.55 eV. MR-ADC(1) qualitatively reproduces the shape of the FCI curve locating an avoided crossing at 15.25 $a_0$ with $\Delta E$ = 0.12 eV. Although at each geometry the MR-ADC(1) error in $\Delta E$ is large (0.3 to 0.5 eV), relative to FCI, the MR-ADC(1) curve is quite parallel to FCI for the short ($<$ 10 $a_0$) and long ($>$ 16 $a_0$) bond distances. In addition, MR-ADC(1) demonstrates a significant improvement over the reference CASCI/CASSCF results, which do not exhibit an avoided crossing at any geometry. The MC-TDA curve is quite close to MR-ADC(1) for the short bond distances ($<$ 14 $a_0$), but significantly deviates from MR-ADC(1) at the longer distances ($>$ 16 $a_0$), showing larger non-parallelity errors relative to FCI. The best results are shown by the sc-QD-NEVPT2 method, which locates an avoided crossing at 11.5 $a_0$ with $\Delta E$ = 0.09 eV, indicating the importance of the two-electron dynamic correlation effects that are not included in MR-ADC(1).

\subsection{Doubly excited states in \ce{C2}}

Finally, we consider a challenging example of the \ce{C2} molecule, whose excited states require very accurate description of static and dynamic correlation.\cite{Roos:1987p399,Bauschlicher:1987p1919,Watts:199p6073,Abrams:2004p9211,Wouters:2014p1501,Holmes:2017p164111} \cref{tab:c2} reports the MR-ADC(1) vertical excitation energies for \ce{C2} at near equilibrium bond distance (\mbox{$r$ = 2.4 $a_0$}) computed using the cc-pVDZ basis set.\cite{Dunning:1989p1007} In addition, we report results from the single-reference Tamm-Dancoff approximation (TDA), CASCI computed using the ground-state CASSCF orbitals (CASCI/CASSCF), and strongly-contracted NEVPT2 (sc-NEVPT2). For all active-space methods, the CASSCF reference wavefunction was computed using the \mbox{(8e, 8o)} active space. As a benchmark, we employ the accurate density matrix renormalization group (DMRG) results obtained by Wouters {\it et al}.\cite{Wouters:2014p1501} 

All low-lying electronic states of \ce{C2} considered in \cref{tab:c2} have a significant multi-reference character. This is evidenced by the TDA method, which incorrectly predicts the ground state and completely misses four out of nine excited states, due to their doubly excited nature. On the contrary, the MR-ADC(1) method correctly assigns the ground state to be \(X\, {}^1\Sigma_g^+\) and provides excitation energies for the doubly excited states. However, for most of the electronic states, the MR-ADC(1) energies significantly overestimate the DMRG results and are close to CASCI/CASSCF. Although for some states (e.g., $B\, {}^1\Delta_g$ or $b\, {}^3\Sigma_g^-$) the MR-ADC(1) and CASCI/CASSCF results are virtually identical, for other states MR-ADC(1) predicts somewhat lower excitation energies, reducing the error relative to DMRG by up to 0.35 eV, with an exception of the $1\, {}^3\Delta_u$ state where a substantial improvement of 0.86 eV is observed. The large errors of MR-ADC(1) are due to the missing description of the two-electron dynamic correlation between core, active, and external orbitals, which is very important in the excited states of \ce{C2}. This is evidenced by the results of the sc-NEVPT2 method, which are in a much closer agreement with DMRG (\mae = 0.21, \std = 0.10 eV) than MR-ADC(1) (\mae = 1.03, \std = 0.69 eV). The two-electron dynamic correlation effects will be incorporated in the second-order MR-ADC approximation (MR-ADC(2)) and are expected to significantly lower the excitation energies, reducing the errors relative to reference values. 

\section{Conclusions}
\label{sec:conclusions}

In this work, we considered a multi-reference formulation of the algebraic diagrammatic construction theory (MR-ADC) for excited states of strongly correlated systems. The MR-ADC approach is an alternative to multi-reference perturbation theories and multi-reference propagator methods and has several attractive properties: (i) it allows for including and systematically improving the description of the dynamic correlation effects outside of the active space, (ii) it describes electronic transitions involving all orbitals (i.e., core, active, and external), (iii) it is based on the Hermitian eigenvalue problem and thus ensures that the excitation energies have real values, (iv) it enables efficient computation of spectroscopic properties (such as transition amplitudes and spectral densities), and (v) allows for simulations of various spectroscopic processes (e.g., valence or core excitations, photoionization). In contrast to the original (single-reference) ADC theory, MR-ADC is more reliable in situations where the multi-reference effects are important in the ground or excited electronic states. Our formulation of MR-ADC is based one the effective Liouvillean formalism of Mukherjee and Kutzelnigg, originally developed for the single-determinant reference wavefunctions.\cite{Mukherjee:1989p257} By generalizing this formalism to multi-determinant states, we arrived at the MR-ADC formulation, which naturally reduces to the conventional ADC theory in the single-reference limit. We performed a perturbative analysis of MR-ADC and outlined a procedure for constructing MR-ADC approximations at each order in perturbation theory.

As a first step in defining the hierarchy of the MR-ADC methods, we presented an implementation of the first-order MR-ADC approximation for the polarization propagator (MR-ADC(1)) and benchmarked its results for two non-interacting water molecules, excitation energies of the Be atom, an avoided crossing in \ce{LiF}, and doubly excited states in \ce{C2}. For the water molecules, we showed that the MR-ADC(1) excitation energies exhibit small size-consistency errors, which originate due to removing linear dependencies in the overlap metric. For the Be atom, we demonstrated that the MR-ADC(1) results converge to the full configuration interaction limit with increasing active space size. In a study of \ce{LiF}, we showed that MR-ADC(1) qualitatively correctly describes an avoided crossing due to the mixing of the ionic and covalent configurations. For \ce{C2}, MR-ADC(1) predicts excitation energies of the doubly excited states, but shows large errors relative to reference values, missing the description of the two-electron dynamic correlation outside of the active space. These correlation effects are incorporated in the second-order MR-ADC approximation and are expected to significantly improve the results. 

We envision many possible extensions of the current work. An immediate extension is the development of the second-order MR-ADC(2) approximation for the polarization propagator, which will incorporate the description of the two-electron dynamic correlation effects that are essential for accurate predictions of the excitation energies and excited-state properties. A further direction is to develop the MR-ADC methods for simulations of other spectroscopic properties, such as photoelectron, X-ray absorption or two-photon absorption spectra. Although the conventional ADC theory has been applied to these problems,\cite{Schirmer:1983p1237,Barth:1985p867,Knippenberg:2012p064107} MR-ADC is expected to be more reliable for systems with challenging electronic structure, such as open-shell molecules and transition metal complexes. Additionally, for systems containing heavy elements, it will be important to combine MR-ADC with the description of relativistic effects. Finally, MR-ADC can be combined with density matrix renormalization group or selected configuration interaction approaches,\cite{Huron:1973p5745,Chan:2011p465} which will enable simulations of spectroscopic properties of multi-reference systems with a large number of strongly correlated electrons. Work along these directions is ongoing in our group.

\section{Acknowledgements}
This work was supported by the start-up funds provided by the Ohio State University.
The author would like to thank Francesco Evangelista for comments on this manuscript and insightful discussions.

\section{Appendix: MR-ADC(1) amplitude equations}
\label{sec:appendix}

Here, we describe how to solve the MR-ADC(1) amplitude equations. The general form of the equations is shown in \cref{eq:proj_amplitude_equations_1,eq:proj_amplitude_equations_2}. The action of the first-order effective Hamiltonian on the reference state can be written as:
\begin{align}
	\label{eq:H1eff_psi0}
	\tilde{H}^{(1)}\ket{\Psi_0} = V\ket{\Psi_0} + (H^{(0)} - E_0) T^{(1)} \ket{\Psi_0} 
\end{align}
where $T^{(1)} = T_1^{(1)} + T_2^{(1)}$ and $E_0$ is the ground-state CASCI/CASSCF energy. 

\subsection{Core-external amplitudes}
First, we consider equations for the core-external $t_{i}^{a(1)}$ and $t_{ix}^{ay(1)}$ amplitudes. For brevity, we omit the perturbation order in our notation, e.g.\@ $t_{i}^{a(1)} \equiv t_{i}^{a}$. Projecting \cref{eq:H1eff_psi0} by $a^{a}_{i}\ket{\Psi_0}$ and $a^{ay}_{ix}\ket{\Psi_0}$ on the left, the amplitude equations can be written in a tensor form:
\begin{align}
	\label{eq:t_amp_0p}
	\mathbf{K^{[0']}} \mathbf{T^{[0']}} = \mathbf{V^{[0']}}
\end{align}
where $[0']$ is a standard notation for the core-external single excitations used in $N$-electron valence perturbation theory (NEVPT)\cite{Angeli:2001p10252,Angeli:2001p297,Angeli:2004p4043} and the tensors are defined as:
\begin{align}
	\mathbf{K^{[0']}} =
	\begin{pmatrix} 
	\bra{\Psi_0}a^{i}_{a} [H^{(0)}, a^{b}_{j}] \ket{\Psi_0} & \bra{\Psi_0}a^{i}_{a} [H^{(0)}, a^{bw}_{jz}] \ket{\Psi_0} \\
	\bra{\Psi_0} a^{ix}_{ay} [H^{(0)}, a^{b}_{j}] \ket{\Psi_0} & \bra{\Psi_0} a^{ix}_{ay} [H^{(0)}, a^{bw}_{jz}] \ket{\Psi_0} 
	\end{pmatrix}
\end{align}
\begin{align}
	\mathbf{T^{[0']}} = 
	\begin{pmatrix} 
	\t{j}{b} \\
	\t{jz}{bw}
	\end{pmatrix}
\end{align}
\begin{align}
	\mathbf{V^{[0']}} = -
	\begin{pmatrix} 
	 \bra{\Psi_0}a^{i}_{a} V\ket{\Psi_0} \\
	 \bra{\Psi_0} a^{ix}_{ay} V\ket{\Psi_0} 
	\end{pmatrix}
\end{align}
The elements of $\mathbf{K^{[0']}}$ reduce to
\begin{align}
	\label{eq:K_0p_1}
	\bra{\Psi_0}a^{i}_{a} [H^{(0)}, a^{b}_{j}] \ket{\Psi_0} &= \d{i}{j} \d{a}{b} (\e{a} - \e{i})  \\
	\label{eq:K_0p_2}
	\bra{\Psi_0}a^{i}_{a} [H^{(0)}, a^{bw}_{jz}] \ket{\Psi_0} &= \d{i}{j} \d{a}{b} (\e{a} - \e{i}) \braket{\Psi_0|a^{w}_{z}|\Psi_0} \\
	\label{eq:K_0p_3}
	\bra{\Psi_0} a^{ix}_{ay} [H^{(0)}, a^{b}_{j}] \ket{\Psi_0} &= \d{i}{j} \d{a}{b} (\e{a} - \e{i}) \braket{\Psi_0|a^{x}_{y}|\Psi_0} \\
	\label{eq:K_0p_4}
	\bra{\Psi_0} a^{ix}_{ay} [H^{(0)}, a^{bw}_{jz}] \ket{\Psi_0} &= \d{i}{j} \d{a}{b} \left[(\e{a} - \e{i}) \braket{\Psi_0|a^{x}_{y}a^{w}_{z}|\Psi_0} \right. \notag \\
	&\left. + \braket{\Psi_0|a^{x}_{y}[H_{act},a^{w}_{z}]|\Psi_0}\right]
\end{align}
For the elements of $\mathbf{V^{[0']}}$, we obtain:
\begin{align}
	 \bra{\Psi_0}a^{i}_{a} V\ket{\Psi_0} &= \h{a}{i} + \sum_{j} \v{aj}{ij} + \sum_{zw} \v{aw}{iz} \braket{\Psi_0|a^{w}_{z}|\Psi_0} \\
	 \bra{\Psi_0} a^{ix}_{ay} V\ket{\Psi_0} &= \h{a}{i} \braket{\Psi_0|a^{x}_{y}|\Psi_0} + \sum_{j} \v{aj}{ij} \braket{\Psi_0|a^{x}_{y}|\Psi_0} \notag \\
	 &+ \sum_{zw} \v{aw}{iz} \braket{\Psi_0|a^{x}_{y}a^{w}_{z}|\Psi_0} 
\end{align}

To solve \cref{eq:t_amp_0p}, we first consider the generalized eigenvalue problem for the matrix $\mathbf{K^{[0']}}$:
\begin{align}
	\label{eq:eig_problem_0p}
	\mathbf{K^{[0']}} \mathbf{X^{[0']}} = \mathbf{S^{[0']}} \mathbf{X^{[0']}} \boldsymbol{\epsilon^{[0']}} 
\end{align}
where the overlap metric has a general form
\begin{align}
	\label{eq:S_0p}
	\mathbf{S^{[0']}} =
	\begin{pmatrix} 
	\bra{\Psi_0}a^{i}_{a} a^{b}_{j} \ket{\Psi_0} & \bra{\Psi_0}a^{i}_{a} a^{bw}_{jz} \ket{\Psi_0} \\
	\bra{\Psi_0} a^{ix}_{ay} a^{b}_{j} \ket{\Psi_0} & \bra{\Psi_0} a^{ix}_{ay} a^{bw}_{jz} \ket{\Psi_0} 
	\end{pmatrix}
\end{align}
Defining $\mathbf{\tilde{X}^{[0']}} = (\mathbf{S^{[0']}})^{1/2} \mathbf{X^{[0']}}$, the $\mathbf{K^{[0']}}$ matrix in \cref{eq:eig_problem_0p} can be expressed as:
\begin{align}
	\mathbf{K^{[0']}} &=  (\mathbf{S^{[0']}})^{1/2} \mathbf{\tilde{X}^{[0']}} \boldsymbol{\epsilon^{[0']}} \mathbf{\tilde{X}^{[0']\dag}} (\mathbf{S^{[0']}})^{1/2}
\end{align}
This allows us to obtain expression for the amplitudes $\mathbf{T^{[0']}}$ in terms of the eigenvalues $\boldsymbol{\epsilon^{[0']}}$, eigenvectors $\mathbf{\tilde{X}^{[0']}}$ and the overlap matrices:
\begin{align}
	\mathbf{T^{[0']}} = (\mathbf{S^{[0']}})^{-1/2} \mathbf{\tilde{X}^{[0']}} (\boldsymbol{\epsilon^{[0']}})^{-1} \mathbf{\tilde{X}^{[0']\dag}} (\mathbf{S^{[0']}})^{-1/2} \mathbf{V^{[0']}}
\end{align}
Solving the eigenvalue problem \eqref{eq:eig_problem_0p} requires diagonalizing the overlap metric $\mathbf{S^{[0']}}$ and projecting out the eigenvectors corresponding to small eigenvalues (see \cref{sec:comp_details} for details). Since the $\mathbf{K^{[0']}}$ matrix elements in \cref{eq:K_0p_1,eq:K_0p_2,eq:K_0p_3,eq:K_0p_4} are zero for $i\ne j$ or $a\ne b$, \cref{eq:eig_problem_0p} can be solved for each block with $i=j$ and $a=b$ independently, which greatly reduces the cost of computing the amplitudes. Although in this work we only deal with the first-order amplitudes, we note that in MR-ADC the extraction of linearly independent amplitudes has to be done separately at each order of perturbation theory.

\subsection{Avoiding computation of 4-RDM for the core-active and active-external amplitudes}
Next, we consider the core-active amplitudes $t_{i}^{x}$ and $t_{ix}^{yw}$ belonging to the $[+1']$ excitation class of NEVPT.
The amplitude equations can be obtained by projecting \cref{eq:H1eff_psi0} by the $a^{x}_{i}\ket{\Psi_0}$ and $a^{xw}_{iz}\ket{\Psi_0}$ configurations and can be written in the form similar to \cref{eq:t_amp_0p}: $\mathbf{K^{[-1']}} \mathbf{T^{[-1']}} = \mathbf{V^{[-1']}}$. As previously, this system of linear equations can be solved by diagonalizing and inverting the matrix $\mathbf{K^{[-1']}}$. However, evaluation of $\mathbf{K^{[-1']}}$ requires the four-particle reduced density matrix (4-RDM), which has a steep $\mathcal{O}(N_{\mathrm{det}} \times N^8_{\mathrm{act}})$ computational scaling with the number of active orbitals $N_{\mathrm{act}}$ and the dimension of the active-space Hilbert space $N_{\mathrm{det}}$. 

Here, we present a different algorithm, which does not require computation of 4-RDM. We start by multiplying both sides of \cref{eq:H1eff_psi0} by \mbox{$(H^{(0)} - E_0)^{-1}a^{x}_{i}\ket{\Psi_0}$} and \mbox{$(H^{(0)} - E_0)^{-1}a^{xw}_{iz}\ket{\Psi_0}$} and set the corresponding projections to zero. For example, for the $a^{x}_{i}\ket{\Psi_0}$ projection we obtain:
\begin{align}
	\label{eq:H1eff_psi0_2}
	\bra{\Psi_0}  a^{i}_{x} T^{(1)} \ket{\Psi_0} &= - \bra{\Psi_0} a^{i}_{x} (H^{(0)} - E_0)^{-1} V\ket{\Psi_0} \notag \\
	&= - \int_0^\infty \bra{\Psi_0} a^{i}_{x} e^{-(H^{(0)} - E_0)\tau} V\ket{\Psi_0} \mathrm{d} \tau
\end{align}
where we expressed the operator resolvent $(H^{(0)} - E_0)^{-1}$ using a Laplace transform as an integral over imaginary time $\tau$.\cite{Sokolov:2016p064102} The resulting amplitude equations can be written in the matrix form:
\begin{align}
	\label{eq:t_amp_+1p}
	\mathbf{T^{[+1']}} = (\mathbf{S^{[+1']}})^{-1} \mathbf{W^{[+1']}}
\end{align}
where the tensor product is carried out over the unique sets of amplitudes only (i.e., $\t{jv}{yu}$, $y>u$) and the tensors are defined as:
\begin{align}
	\mathbf{T^{[+1']}} = 
	\begin{pmatrix} 
	\t{j}{y} \\
	\t{jv}{yu}
	\end{pmatrix}
\end{align}
\begin{align}
	\label{eq:S_+1p}
	\mathbf{S^{[+1']}} =
	\begin{pmatrix} 
	\bra{\Psi_0} a^{i}_{x} a^{y}_{j} \ket{\Psi_0} & \bra{\Psi_0} a^{i}_{x} a^{yu}_{jv} \ket{\Psi_0} \\
	\bra{\Psi_0} a^{iz}_{xw} a^{y}_{j} \ket{\Psi_0} & \bra{\Psi_0} a^{iz}_{xw} a^{yu}_{jv} \ket{\Psi_0} 
	\end{pmatrix}
\end{align}
\begin{align}
	\mathbf{W^{[+1']}} 
	&= -
	\begin{pmatrix} 
	 \int_0^\infty \bra{\Psi_0} a^{i}_{x} e^{-(H^{(0)} - E_0)\tau} V\ket{\Psi_0} \mathrm{d} \tau \\
	 \int_0^\infty \bra{\Psi_0} a^{iz}_{xw} e^{-(H^{(0)} - E_0)\tau} V\ket{\Psi_0} \mathrm{d} \tau
	\end{pmatrix} \notag \\
	\label{eq:W_p1}
	&= -
	\begin{pmatrix} 
	\int_0^\infty e^{\e{i} \tau} \bra{i(\tau)} \c{x} \ket{\Psi_0} \mathrm{d} \tau \\
	\int_0^\infty e^{\e{i} \tau} \bra{i(\tau)} \c{x} \c{w} \a{z}\ket{\Psi_0} \mathrm{d} \tau
	\end{pmatrix}
\end{align}
The matrix elements of $\mathbf{W^{[+1']}}$ can be evaluated by the imaginary-time propagation of the active-space state $\ket{i(\tau)} = e^{-(H^{(0)} - E_0)\tau}\ket{i}$, where
\begin{align}
	\ket{i} &\equiv \sum_{y} \h{y}{i} \c{y} \ket{\Psi_0} + \sum_{yk} \v{yk}{ik} \c{y} \ket{\Psi_0} \notag \\
	&+ \frac{1}{2} \sum_{wzy} \v{yw}{iz} \c{y} \c{w} \a{z} \ket{\Psi_0}
\end{align}
and integrating \cref{eq:W_p1} numerically over a set of time steps $\tau_k$. The details of the imaginary-time propagation algorithm can be found in \cref{sec:comp_details} and in Ref.\@ \citenum{Sokolov:2016p064102}. Once $\mathbf{W^{[+1']}}$ is determined, we diagonalize the overlap metric $\mathbf{S^{[+1']}}$, project out the linearly dependent eigenvectors with small eigenvalues, and compute the amplitudes $\mathbf{T^{[+1']}}$ according to \cref{eq:t_amp_+1p}. 

For the active-external excitations $t_{x}^{a}$ and $t_{xy}^{aw}$ corresponding to the $[-1']$ excitation class of NEVPT, the amplitude equations have a similar form:
\begin{align}
	\mathbf{T^{[-1']}} = (\mathbf{S^{[-1']}})^{-1} \mathbf{W^{[-1']}}
\end{align}
\begin{align}
	\mathbf{T^{[-1']}} = 
	\begin{pmatrix} 
	\t{y}{b} \\
	\t{yv}{bu}
	\end{pmatrix}
\end{align}
\begin{align}
	\label{eq:S_-1p}
	\mathbf{S^{[-1']}} =
	\begin{pmatrix}
	\bra{\Psi_0} a^{x}_{a} a^{b}_{y} \ket{\Psi_0} & \bra{\Psi_0} a^{x}_{a} a^{bu}_{yv} \ket{\Psi_0} \\
	\bra{\Psi_0} a^{xz}_{aw} a^{b}_{y} \ket{\Psi_0} & \bra{\Psi_0} a^{xz}_{aw} a^{bu}_{yv} \ket{\Psi_0} 
	\end{pmatrix}
\end{align}
\begin{align}
	\mathbf{W^{[-1']}} 
	&= -
	\begin{pmatrix} 
	\int_0^\infty e^{-\e{a} \tau} \bra{a(\tau)} \a{x} \ket{\Psi_0} \mathrm{d} \tau \\
	\int_0^\infty e^{-\e{a} \tau} \bra{a(\tau)} \c{w} \a{z} \a{x}\ket{\Psi_0} \mathrm{d} \tau
	\end{pmatrix}
\end{align}
\begin{align}
	\ket{a} &\equiv \sum_{y} \h{a}{y} \a{y} \ket{\Psi_0} + \sum_{yk} \v{ak}{yk} \a{y} \ket{\Psi_0} \notag \\
	&+ \frac{1}{2} \sum_{wzy} \v{aw}{yz} \c{w} \a{z} \a{y} \ket{\Psi_0}
\end{align}

Avoiding the evaluation of 4-RDM lowers the computational cost of solving the core-active and active-external amplitude equations from $\mathcal{O}(N_{\mathrm{det}} \times N^8_{\mathrm{act}})$ to \mbox{$\mathcal{O}(N_\tau \times N_{\mathrm{det}} \times N_{\mathrm{core}} \times N^4_{\mathrm{act}})$} and \mbox{$\mathcal{O}(N_\tau \times N_{\mathrm{det}} \times N_{\mathrm{ext}} \times N^4_{\mathrm{act}})$}, respectively, where $N_\tau$ is the number of time steps in imaginary time ($\sim$ 10 to 40), $N_{\mathrm{core}}$ is the number of core orbitals, and $N_{\mathrm{ext}}$ is the number of external orbitals.

\end{document}